\documentclass[%
 reprint,
groupedaddress,
amsmath,
amssymb,
aps,
prd,
twocolumn,preprintnumbers,floatfix,nofootinbib]{revtex4-2}

\usepackage[pdftex]{graphicx}
\usepackage{dcolumn}
\usepackage{bm}
\usepackage{hyperref}

\newcommand{\be}{\begin{equation}}
\newcommand{\ee}{\end{equation}}
\newcommand{\bea}{\begin{eqnarray}}
\newcommand{\eea}{\end{eqnarray}}

\usepackage{mathrsfs}
\usepackage{amsmath, amsthm, amssymb}
\usepackage{color}
\usepackage{epstopdf}
\usepackage{amssymb}
\usepackage{verbatim}
\usepackage{enumerate}
\usepackage{float}
\usepackage[caption = false]{subfig}
\usepackage[normalem]{ulem}  
\usepackage{epsfig}
\usepackage{cleveref}

\newcommand{\araa}{ARA\&A}		
\newcommand{\apjl}{ApJLett}		
\newcommand{\mnras}{MNRAS}		
\newcommand{\jcap}{JCAP}		

\newcommand{\physrep}{Phys.~Rep.}   

\begin{document}
\title{New insights into the formation and growth of boson stars in dark matter halos}

\author{Jiajun Chen$^1$}
\email{jiajun@astro.physik.uni-goettingen.de}
\author{Xiaolong Du$^2$}
\email{xdu@carnegiescience.edu}
\author{Erik W. Lentz$^1$}
\email{erik.lentz@uni-goettingen.de}
\author{David J. E. Marsh$^1$}
\email{david.marsh@uni-goettingen.de}
\author{Jens C. Niemeyer$^1$}
\email{niemeyer@astro.physik.uni-goettingen.de}
\affiliation{
$^1$Institut f\"ur Astrophysik, Georg-August-Universit\"at G\"ottingen, Friedrich-Hund-Platz 1, D-37077 G\"ottingen, Germany\\
$^2$Carnegie Observatories, 813 Santa Barbara Street, Pasadena, CA 91101, U.S.A}

\date{\today}

\begin{abstract}
This work studies the formation and growth of boson stars and their surrounding miniclusters by gravitational condensation using non-linear dynamical numerical methods. Fully dynamical attractive and repulsive self-interactions are also considered for the first time. In the case of pure gravity, we numerically prove that the growth of boson stars inside halos slows down and saturates as has been previously conjectured, and detail its conditions. Self-interactions are included using the Gross-Pitaevskii-Poisson equations. We find that in the case of strong attractive self-interactions the boson stars can become unstable and collapse, in agreement with previous stationary computations. At even stronger coupling, the condensate fragments. Repulsive self-interactions, as expected, promote boson star formation, and lead to solutions with larger radii.
\end{abstract}

\maketitle

\section{Introduction}\label{sec:intro}
Dark matter(DM) is a hypothetical form of matter, which makes up nearly $27\%$ of the contents in our Universe \cite{Aghanim:2018eyx}.
A popular idea is that the dark matter could be formed of light (pseudo-)scalar particles with large occupation number so that they can be described by a classical scalar field $\phi$, see e.g. \cite{Dine:1982ah,Suarez:2013iw,Preskill:1982cy,Abbott:1982af,Guth_2015,Widrow:1993qq,Uhlemann:2014npa}. 
The potential of the scalar field can be expanded for small field values as \cite{Sikivie:2006ni,Arvanitaki:2009fg,2017PhRvL.118a1301L}
\begin{equation}
V(\phi) = \frac{1}{2} m^2 \phi^2 + \frac{1}{2}m^2 g \phi^4+...,
\end{equation}
where $m$ is the particle mass, $g$ is a dimensional self-interaction coupling constant (see below), and natural units, $\hbar=c=1$, are used. Depending on the detailed models, the particle's mass and coupling constant have different values. For instance, QCD axions have masses in the range $10^{-11}-10^{-2}$ eV \cite{pecceiquinn1977,weinberg1978,wilczek1978,Kim:1979if,Shifman:1979if,Dine:1982ah,Zhitnitsky:1980tq,1981PhLB..104..199D,2015PhRvD..91h4011A,2015JCAP...02..006P,Marsh:2015xka,Tanabashi:2018oca,luzio2020landscape}, while the ultralight boson particles have masses as low as $10^{-22}-10^{-19}$ eV \cite{1990PhRvL..64.1084P,2000PhRvD..62j3517S,hu2000,2000ApJ...534L.127P}.
These different parameters can influence gravitational structure formation and the mass distribution of scalar field dark matter in our Universe.
For example, for ultralight bosons without self-interaction, their dynamics can be computed in the non-relativistic regime using the Schr\"{o}dinger-Poisson (SP) equations. For bosons with repulsive or attractive self-interactions, they have a non-zero self-coupling constant, leading to the Gross-Pitaevskii-Poisson (GPP) equations. Solving the SP or GPP equations, we know that the finite energy ground state solution (for weak self-interactions) of these systems is a soliton: a localized lump of boson energy density held together by the competing forces of gravity, self-interactions, and gradient energy~\cite{2015MNRAS.451.2479M, Hui:2016ltb, PhysRevD.42.384,Widrow:1993qq}. Such solitonic solutions are also known as boson stars~\cite{1968PhRv..172.1331K,1969PhRv..187.1767R,1994PhRvL..72.2516S,Alcubierre:2003sx,2017JCAP...03..055H,2017PhRvL.118a1301L,Liddle:1993ha}.

The formation of boson stars has been observed in numerical simulations in a variety of situations relevant to dark matter structure formation~\cite{Schive:2014dra, Schive:2014hza, Schwabe:2016rze, Levkov:2018kau, Widdicombe:2018oeo, Eggemeier:2019jsu}. The growth rate of boson stars is an important quantity since the more massive and dense stars are easier to observe. Ultimately, if we can understand the masses and abundances of boson stars, then we can predict their observational signatures, in instruments such as haloscopes or gravitational waves detectors, or by gravitational lensing, or via their decay products~\cite{2016PhRvL.116t1301B,Wu:1998ju,Hinshaw:2008kr,Natarajan:2017sbo,2003ARA&A..41..645R,2005MPLA...20.1021B,2010arXiv1001.3706M,1988PhLB..214..403E}. 

For bosons without self-interaction, boson stars can form due to gravitational condensation from isotropic initial conditions \cite{Levkov:2018kau}. After nucleation, boson stars start to acquire mass from the surrounding field, with an initial mass growth rate $\propto t^{1/2}$, as obtained by \cite{Levkov:2018kau}. However, due to computational limitations, the end-stage evolution of boson stars has yet to be observed despite predictions that saturation of the mass growth rate will drop to $\propto t^{1/8}$~\cite{Eggemeier:2019jsu}. In fact, if saturation is reached within a Hubble time, the end-stage evolution of boson stars is the main factor determining their mass distribution at present. Thus studying the end-stage evolution of boson stars can help us to search for, and possibly observe, boson stars in the Universe. 

Attractive or repulsive self-interaction can further influence the evolution of the boson systems. Dynamical boson star formation in this case has not been studied using non-linear numerical methods. One example is for QCD axions, where the attractive self-interaction could lead to the collapse of boson stars at a critical mass~\cite{Levkov_2017}. Therefore, we need to study the evolution of these kinds of bosons in order to understand mass distribution in our Universe. 

Using pseudo-spectral methods~\cite{Fornberg1987}, we study the evolution of systems with different scalar potentials. We find the following results in our simulation:
\begin{itemize}
\item  For ultralight bosons without self-interaction, the saturation of boson stars occurs in miniclusters. The mass growth rate of boson stars drops from $\propto t^{1/2}$ to $\propto t^{1/8}$
as conjectured by~\cite{Eggemeier:2019jsu}.
\item For bosons with attractive self-interaction, such as QCD axions ($10^{-20} {\rm GeV}^{-2}<g<10^{-35}{\rm GeV}^{-2}$), the self-interaction can cause collapse of boson stars above a critical mass. However, it does not affect condensation and early-stage evolution of boson stars in miniclusters.
\item For bosons with a repulsive self-interaction, condensation and growth of boson stars is promoted. At strong coupling, the resulting boson stars are well described by the Thomas-Fermi profile~\cite{Maga_a_2012}, with a larger radius than the case with no self-interaction.
\item For strong attractive self-interactions, the condensate can fragment and form multiple boson stars even in a small simulation box (see also \cite{Amin_2019} for the case with a saturated scalar potential).
\end{itemize}
Each of the above results are novel, and have not been found before in the dynamical and non-linear regime. 
The observed growth rate of boson stars in our simulations has implications for the expected astrophysical abundance and mass function of boson stars in both the Fuzzy DM mass range $m\sim 10^{-22}\rightarrow 10^{-19}\text{ eV}$, and the QCD axion mass range, $m\sim 10^{-10}\rightarrow 10^{-2}\text{ eV}$. The boson star mass function of fuzzy DM has implications for the cusp-core problem of dwarf spheroidal galaxies (e.g. Ref.~\cite{2015MNRAS.451.2479M}), the dynamics of old star clusters (e.g. Refs.~\cite{Marsh:2018zyw}), the rotation curves of low surface brightness galaxies (e.g. Ref.~\cite{Bar:2018acw}), the kinematics in the Milky Way centre~\cite{2020ApJ...889...88L}, and possibly for the formation of supermassive black holes~\cite{Schive:2014hza}. For the QCD axion, the boson star mass function has implications for radio astronomy and other indirect dark matter detection methods~\cite{Hook:2018iia}, in addition to direct detection methods~\cite{JacksonKimball:2017qgk}. More generally, a saturation mass of boson stars has implications for gravitational wave searches for exotic compact objects~\cite{Giudice:2016zpa}, where the saturation mass implies a maximum compactness for boson stars formed by gravitational condensation and accretion. 

We start in Sec.~\ref{GPP_Equations} with introducing the GPP equations and our initial distributions. In Sec.~\ref{Condensation_Saturation_boson_Star} we study the formation and saturation of boson stars for bosons without self-interaction. In Sec.~\ref{Self_Inter_boson_Star} and Sec.~\ref{boson_Repulsive_Self_Inter} we study the evolution of bosons with attractive self-interaction and repulsive self-interaction, respectively. In Sec.\ref{Powerful_Attractive_Self_Inter}, we study the condensation of multiple fragments.
Finally, we present our conclusions in Sec.~\ref{Conclusion}. In Appendix, we discuss the pseudo-spectral method, convergence analysis, soliton solutions to the GPP equations, condensation time for gravitational and self-interactions.

\section{The Gross-Pitaevskii-Poisson equations and Initial distributions}
\label{GPP_Equations}
In the non-relativistic, low-density and low-velocity limits, we can rewrite the scalar field $\phi$ as
\begin{equation}
\phi= \sqrt{\frac{2}{m}} {\rm Re}(\psi e^{-i m t}).
\label{eq:phi_psi}
\end{equation}
The complex wave function $\psi$ at lowest order satisfies the GPP equations
\cite{Chavanis:2011zm,Eby:2015hsq}
\begin{eqnarray}
i\frac{\partial}{\partial{t}}\psi&=&-\frac{1}{2m}\nabla^2\psi + m V\psi+g|\psi|^2\psi,
\label{eq:GPP1}
\\
\nabla^2{V}&=&4 \pi G m\left(|\psi|^2-n\right),
\label{eq:GPP2}
\end{eqnarray}
where $G$ is Newton's gravitational constant, $V$ is the gravitational potential, and $n$ is the mean number density. Eqs. (\ref{eq:GPP1}) and (\ref{eq:GPP2}) can be written in a dimensionless form following the definitions in Ref.~\cite{Levkov:2018kau}: substitutions $x=\widetilde{x}/(m v_0)$, $t=\widetilde{t}/(mv_0^2)$, $V=\widetilde{V} v_0^2$ and $\psi=\widetilde{\psi}v_0^2\sqrt{m/(4 \pi G)}$, $g=\widetilde{g}\,4\pi G/v_0^2$,
where $v_0$ is a reference velocity, e.g. the characteristic velocity of the initial state.
The dimensionless equations are given by
\begin{eqnarray}
i\frac{\partial}{\partial{\widetilde{t}}}\widetilde{\psi}&=&-\frac{1}{2}\widetilde{\nabla}^2\widetilde{\psi} + \widetilde{V}\widetilde{\psi}+\widetilde{g}|\widetilde{\psi}|^2\widetilde{\psi},
\label{eq:GPP1_dim}
\\
\widetilde{\nabla}^2{\widetilde{V}}&=&|\widetilde{\psi}|^2-\widetilde{n},
\label{eq:GPP2_dim}
\end{eqnarray}

We use the wave function momentum distribution~\cite{Levkov:2018kau}, $|\psi_{\vec{p}}|^2=N\delta(|\vec{p}|-mv_0)$ in a periodic box of size $L$ as initial conditions, where $N\equiv n L^3$ is number of non-relativistic bosons in the box. Performing an inverse Fourier transform on $\psi_{\vec{p}} e^{i S}$ with $S$ a random phase, we obtain an isotropic initial distribution in position space, $\psi(\vec{x},0)$. This initial distribution follows from the uncertainty principle: exact knowledge of $\vec{p}$ gives complete uncertainty in $\vec{x}$. In order to study isolated halos/miniclusters, we run simulations in a box of size $\widetilde{L}>2\pi/\widetilde{k}_J$ with $\widetilde{k}_J=(4\widetilde{n})^{1/4}$ the dimensionless Jeans wavenumber, since non-relativistic boson gas forms clumps at scales larger than $2\pi/\widetilde{k}_J$ due to Jeans instability~\cite{khlopov_scalar}. 
To study the influence of self-interactions, we vary the dimensionless coupling constant $\widetilde{g}$ within the range $[-100,100]$.

\section{Condensation of bosons}
\subsection{Condensation of bosons without self-interactions}
\label{Condensation_Saturation_boson_Star}
For bosons without self-interactions ($\widetilde{g} = 0$), the GPP equations can be simplified as SP equations:
\begin{eqnarray}
i\frac{\partial}{\partial{\widetilde{t}}}\widetilde{\psi}&=&-\frac{1}{2}\widetilde{\nabla}^2\widetilde{\psi} + \widetilde{V}\widetilde{\psi},
\label{eq:schro}
\\
\widetilde{\nabla}^2{\widetilde{V}}&=&|\widetilde{\psi}|^2-\widetilde{n}.
\label{eq:poiss}
\end{eqnarray}

It has been observed that gravity leads to the formation of a gravitationally virialized DM halo (for certain cosmologies called a ``minicluster''), and eventually to the condensation of a boson star at its center~\cite{Levkov:2018kau}. The condensation time, $\tau_{\rm gravity}$, can be derived from the theory of relaxation in the SP equations and the Landau equation~\cite{Schive:2014hza, Levkov:2018kau,Kirkpatrick:2020fwd}, which is given in terms of the radius of miniclusters, $R$, characteristic velocity, $v$, and density, $n$, of the minicluster~\cite{Levkov:2018kau}:
\begin{equation}
\tau_{\rm gravity}=\frac{b\sqrt{2}}{12\pi^3}\frac{mv^6}{G^2n^2\Lambda},
\label{eq:condensation_time}
\end{equation}
where $\Lambda = \log(mvR)$ is the Coulomb logarithm, and $b$ is an $\mathcal{O}(1)$ co-efficient to be determined by simulation. After condensation, boson stars have been shown to acquire mass from the surrounding gas of particles, with the subsequent growth rate~\cite{Levkov:2018kau}, 
\begin{equation}
M_{\rm *}(t) \simeq M_{*,0}\left(\frac{t-\tau_{\rm gravity}}{\tau_{\rm gravity}}\right)^{1/2},
\label{eq:boson_star_growth_1}
\end{equation}
where $M_*$ is the mass of the boson star, $M_{*,0}$ is the mass of boson star at $t = \tau_{\rm gravity}$.

The question arises as to whether the growth in Eq.~\eqref{eq:boson_star_growth_1} continues forever or saturates. We know immediately after a boson star has been formed, its growth rate is in accordance with Eq.~\eqref{eq:boson_star_growth_1}.
As this boson star grows, surrounding bosons become gravitationally bound to it in a halo or atmosphere (the minicluster surrounding the star). The halo surrounding the boson star contains granular structure on the scale of the de Broglie wavelength, which can be modelled as consisting of transient ``quasi particles"~\cite{Hui:2016ltb,Schive:2014hza}. As the boson star grows in mass, its radius contracts. At a particular mass, $M_{*, \rm sat}$, the size of the boson star will be of order that of the granular structure. At this time, it has been conjectured that the hot atmosphere will reach virial equilibrium with the star, causing the mass growth to slow down~\cite{Eggemeier:2019jsu}. The transition has been predicted occur at $v_{\rm vir*}{\approx}v_{\rm halo}$~\cite{Eggemeier:2019jsu}, where $v_{\rm vir*}$ and $v_{\rm halo}$ are the viral velocity of the boson star and minicluster respectively. We call this time the saturation time, $\tau_{\rm sat}$. The saturation time is estimated by considering the viral velocity in the gravitational potential of the soliton approximately given by~\cite{Hui:2016ltb}
\begin{equation}
v_{\rm vir*}(M_*){\simeq}\frac{GM_*m}{\hbar}.
\label{eq:v_star}
\end{equation}
Exploiting that $(M_{*, \rm sat}/v_{\rm sat,*})^{3/4} = (mG)^{-3/4}$, and combining this with Eq.~\eqref{eq:boson_star_growth_1} gives
\begin{equation}
M_{\rm *}(t_{\rm star}){\simeq}M_{*, \rm sat}\left(\frac{t_{\rm star}}{\tau_{\rm sat}}\right)^{1/8}
\label{eq:boson_star_growth_2}
\end{equation}
with~\cite{Eggemeier:2019jsu}
\begin{equation}
\tau_{\rm sat}=\frac{b\sqrt{2}}{12\pi^3}\frac{m v_{*,\rm sat}^6}{G^2 n^2 \log(mv_{\rm halo}R_{\rm halo})},
\label{eq:saturation_time}
\end{equation}
where $M_{*, \rm sat}$ is the boson star mass at the saturation time, $v_{*,\rm sat}=v_{\rm vir*}(M_{*, \rm sat})$, $R_{\rm halo}$ is the radius of the halo, and $v_{\rm halo}$ is the virial velocity of the halo.

Due to computational limitations, the prediction of the saturation of boson stars has not been verified~\cite{Levkov:2018kau,Eggemeier:2019jsu}. In the rest of this subsection, by running a large number of numerical simulations past the estimated saturation time $t>\tau_{\rm sat}$, we are able to demonstrate that the growth of boson stars in miniclusters indeed saturates as predicted.

We show the evolution of boson stars from our simulations with different $L$ and $N$ to statistically verify our results. From the simulations, we obtain the change of energy, the growth rate of boson stars, etc.  

\subsubsection{Condensation of boson stars}
\label{Formation_boson_Star}
\begin{figure}[htbp]
\centering
\textbf{Boson star growth with no self-interactions}\par\medskip
\includegraphics[width=\columnwidth]{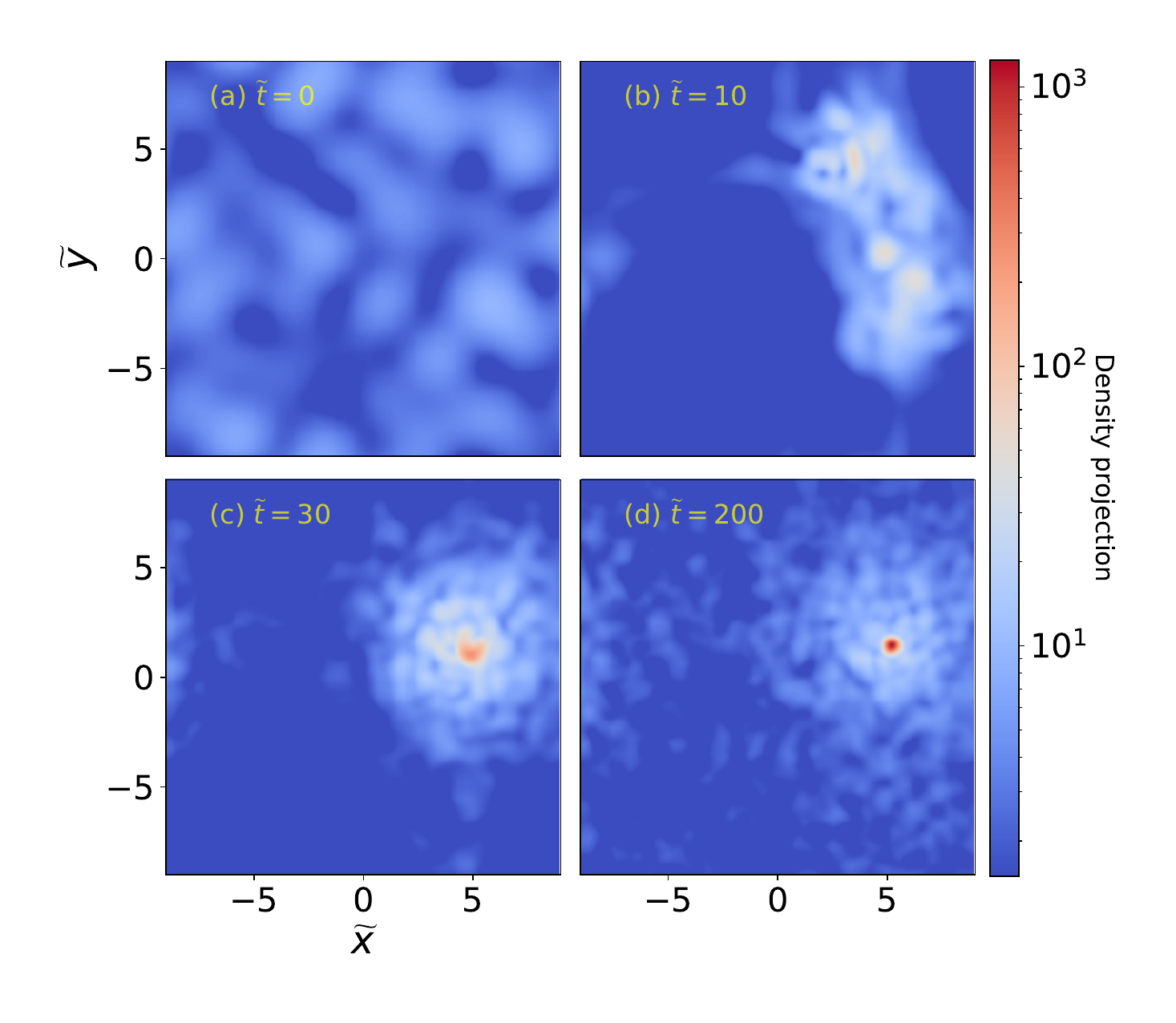}
\caption{Snapshots of the density field from one simulation with $\widetilde{N}=1005.3$, $\widetilde{L}=18$. (a) Projected density at the initial time. (b) Projected density at $\widetilde{t}=10$, which shows that minicluster is forming in the box. (c )Projected density at $\widetilde{t}=30$. (d) Projected density at $\widetilde{t}=200$. A single dense object is visible at the centre of the minicluster.}
\label{fig:1005_18_projection}
\end{figure}

\begin{figure}[htbp]
\centering
\includegraphics[width=\columnwidth]{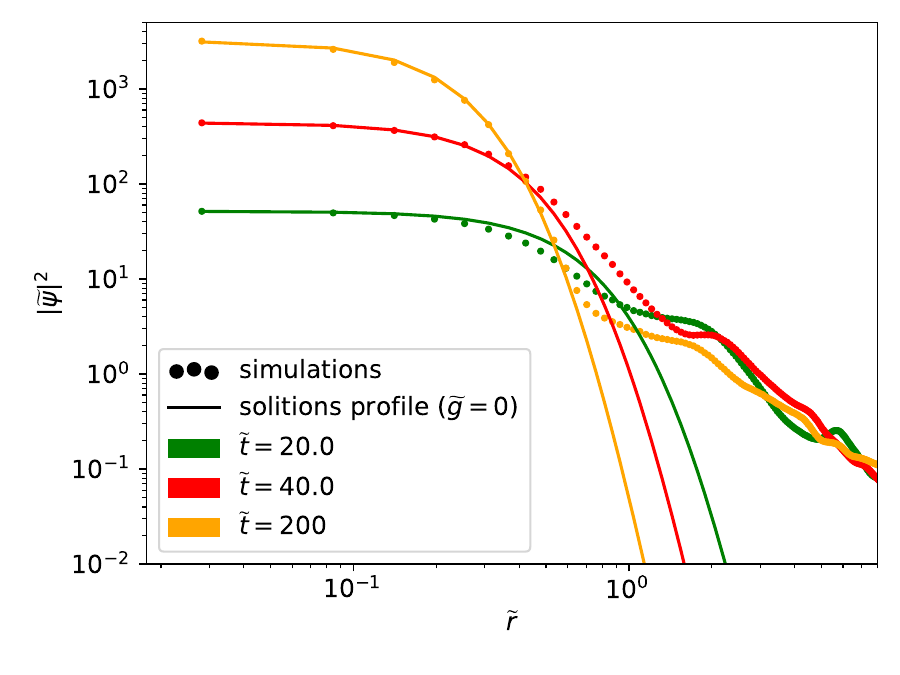}
\caption{Density profiles of the minicluster at different times (colored dots) compared with solitonic profiles (solid lines) as given by Eq.~\eqref{eq:soliton_profile} with the same central densities.}
\label{fig:1005_18_profile}
\end{figure}

Numerically solving the SP equations  at box size $10<\widetilde{L}<30$ and total mass $500<\widetilde{N}<1800$, we observe the formation of a boson star and its surrounding halo/minicluster. One example is shown in Fig.~\ref{fig:1005_18_projection}: box size $\widetilde{L}=18$ and total mass $\widetilde{N}=1005.3$. We can see a minicluster forming gradually from $\widetilde{t} \approx 10$ to $\widetilde{t} \approx 30$. After that, a dense and nearly spherically symmetric object appears and grows in the center of the minicluster. We find that the radial density profile of the minicluster from this most dense point coincides with the density profile of a soliton solution at $0<\widetilde{r}<1$ (soliton density profiles are described in \cref{APPE:soliton}), and a power law at $\widetilde{r}>1$ (see Fig.~\ref{fig:1005_18_profile}).
We also find that there is always one, and only one, boson star formed in each minicluster\footnote{See later for when mutliple boson stars are formed.}. The region outside the boson star has a radial density profile consistent with cold DM on scales larger than the de Broglie wavelength, and with granular structure below it.
These results are fully consistent with results of Refs.~\cite{Schive:2014hza,Levkov:2018kau,Eggemeier:2019jsu}.

\subsubsection{Growth of boson stars}
\label{Growth_boson_Star}
\begin{figure}[htbp]
\centering
\includegraphics[width=\columnwidth]{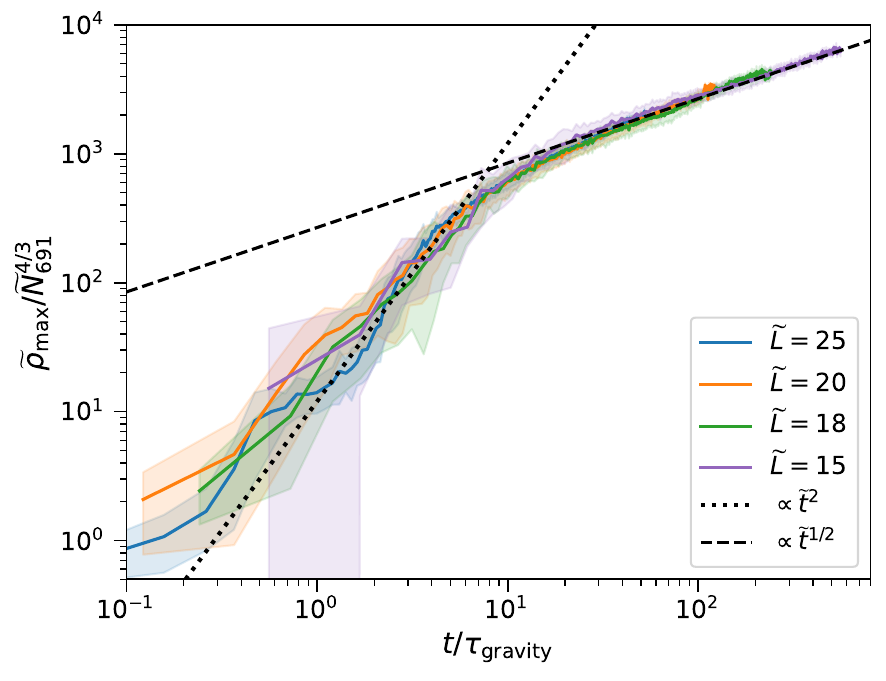}
\caption{The mean stacked maximum density evolution (solid lines)
for different box sizes $\widetilde{L} = 25, 20, 18, 15$ and total mass $\widetilde{N} = 691, 754, 817, 880, 942, 1005, 1131$. The data from simulation with the same box size $\widetilde{L}$ but different total mass $\widetilde{N}$ are divided into 500 time bins. The shaded regions show the $1-\sigma$ intervals. The time and maximum density are normalized by the condensation time, $\tau_{\rm gravity}$ and the total mass, $\widetilde{N}_{691}^{4/3}$, where $\widetilde{N}_{691} = \widetilde{N}/691$. Note that here $\tau_{\rm gravity}$ is computed using Eq. (\ref{eq:condensation_time}) for the initial configuration, i.e. $R=L$, $v=v_0$, and $n=N/L^3$, to avoid ambiguities in the definitions of halo radius and density.}
\label{fig:1005_18_max_density_tog}
\end{figure}

{Fig.~\ref{fig:1005_18_max_density_tog} shows the evolution of mean, normalised, stacked maximum density for our ensemble of simulations. The boson stars form at $t \approx \tau_{\rm gravity}$. After that, we find the growth rate of maximum density $\propto \widetilde{t}^{2}$ at $t \lesssim 10\tau_{\rm gravity}$, falling to $\propto \widetilde{t}^{1/2}$ at the the saturation time. Using the result that for a soliton $M \propto \rho_c^{1/4}$~\cite{Schive:2014hza}, we obtain that the initial mass growth rate of the boson star is $\propto \widetilde{t}^{1/2}$. When the size of the boson star becomes smaller than the granular structure in the surrounding halo, the boson star growth saturates and drops to $\propto \widetilde{t}^{1/8}$ at the transition time, as predicted by Eq.~\eqref{eq:boson_star_growth_2}. Therefore, the saturation of boson star growth indeed occurs in our system, and the asymptotic mass growth rate of the boson star matches the theoretical prediction~\cite{Eggemeier:2019jsu}. Furthermore, we find that during the end stages of evolution, the maximum density can be normalized by $\widetilde{N}^{4/3}$. We believe that the reason is the core-halo mass relation~\cite{Schive:2014hza}, $M_{\rm *} \propto M_{\rm halo}^{1/3}$, where $M_{\rm halo}$ is mass of the halo, and we assume the mass of stable halos in box is proportional to the total mass in the box, $\widetilde{N}$.}

\subsection{Condensation of bosons with self-interactions}
\label{Self_Inter_boson_Star}
Here we include self-interaction. Self-interactions can promote condensation of bosons. Simulating the GPP equations, we study the evolution of bosons with self-interactions.

\subsubsection{Bosons with attractive self-interactions}
\begin{figure}[htbp]
\centering
\textbf{Boson star growth with weak attractive self-interactions}\par\medskip
\includegraphics[width=\columnwidth]{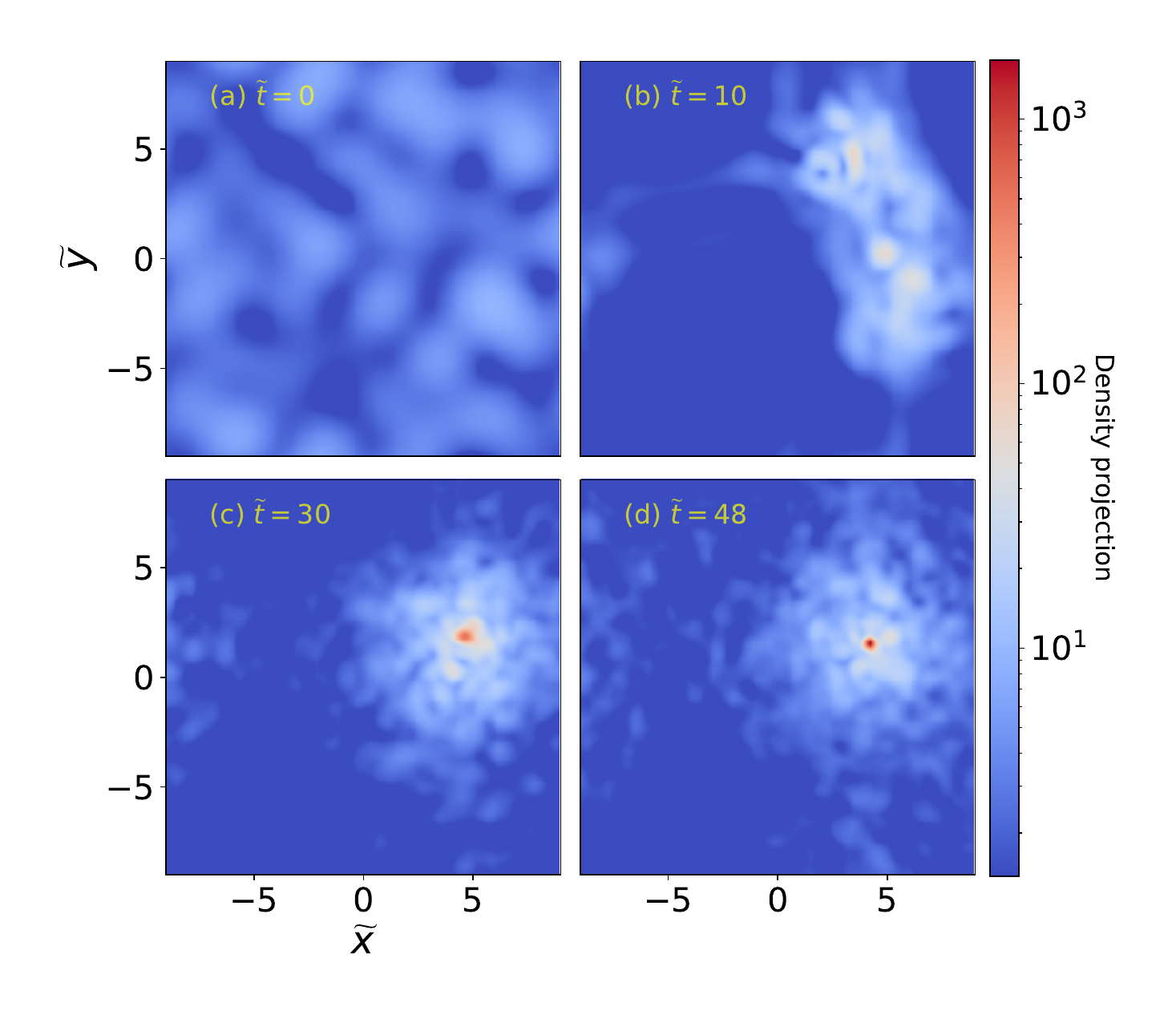}
\caption{Snapshots of the density field from one simulation with $\widetilde{N}=1005.3$, $\widetilde{L}=18$, $\widetilde{g}=-0.007$. (a) Projected density at the initial time. (b) Projected density at $\widetilde{t}=10$, which shows that minicluster is forming in the box. (c) Projected density at $\widetilde{t}=30$. (d) Projected density at $\widetilde{t}=48$. Compared with the case without self-interactions, the boson star formed at the center is denser.}
\label{fig:densityprojection_g=0.007}
\end{figure}

\begin{figure}[htbp]
\centering
\includegraphics[width=\columnwidth]{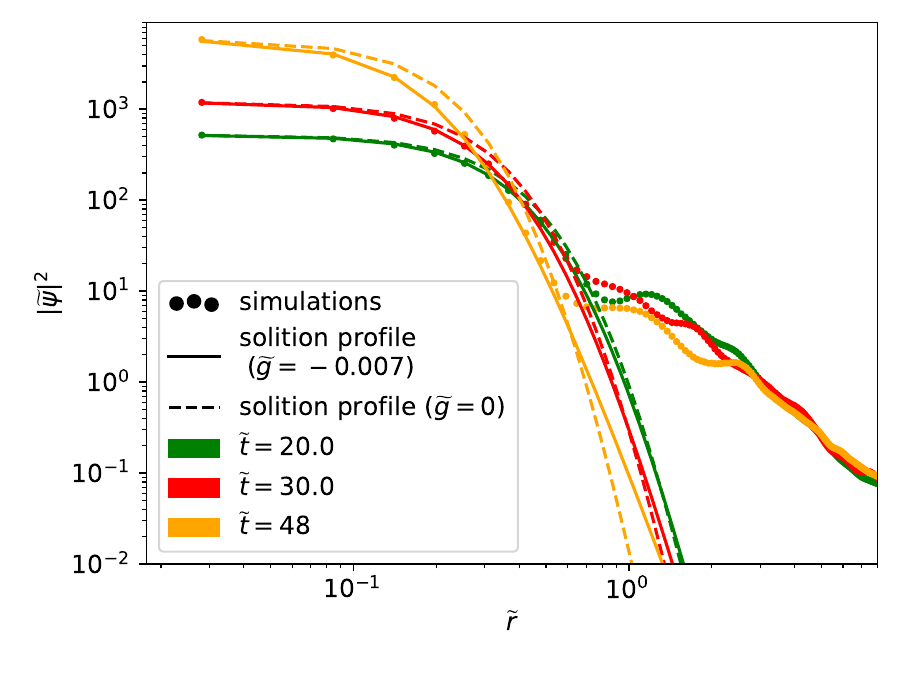}
\caption{Density profiles of the minicluster at different times (colored dots) from simulations of the GPP equations with $\widetilde{g} = -0.007$. Solitonic profiles given by Eq. ~\eqref{eq:sol_rho_gen} and Eq. ~\eqref{eq:soliton_profile} are plotted in solid and dashed lines, respectively.}
\label{fig:densityprofile_g0point007}
\end{figure}

Levkov et al.~\cite{Levkov:2018kau} predict that sufficiently weak attractive self-interactions, like those of the QCD axion, have a negligible effect on boson star formation. However, this prediction has not been directly demonstrated. For bosons with weak attractive self-interaction, such as QCD axions with $v \approx 10^{-9}$, and decay constant $f_a \approx 10^{11}{\rm GeV}$, where $f_a=-1/\sqrt{-12g}$, we obtain an estimate on the self-interaction coupling of $\widetilde{g} \approx -10^{-2}$.
We run some simulations at this range of $\widetilde{g}$. One of these simulations is shown in Fig.~\ref{fig:densityprojection_g=0.007}. We can see the process of formation of the minicluster and condensation of the boson star. This process is similar to the pure gravity case, Fig.~\ref{fig:1005_18_projection}. The radial density profiles of the minicluster and analytic profiles of soliton with and without self-interactions are given in Fig.~\ref{fig:densityprofile_g0point007} and fitted by Eq.~\eqref{eq:sol_rho_gen} and Eq.~\eqref{eq:soliton_profile}, respectively. We discover that the radial density profile of the minicluster coincides with the density profile of a soliton solution at $0<\widetilde{r}<1$, with the case with the correct value of $\widetilde{g}$ providing a better fit. 

\begin{figure}[htbp]
\centering
\includegraphics[width=\columnwidth]{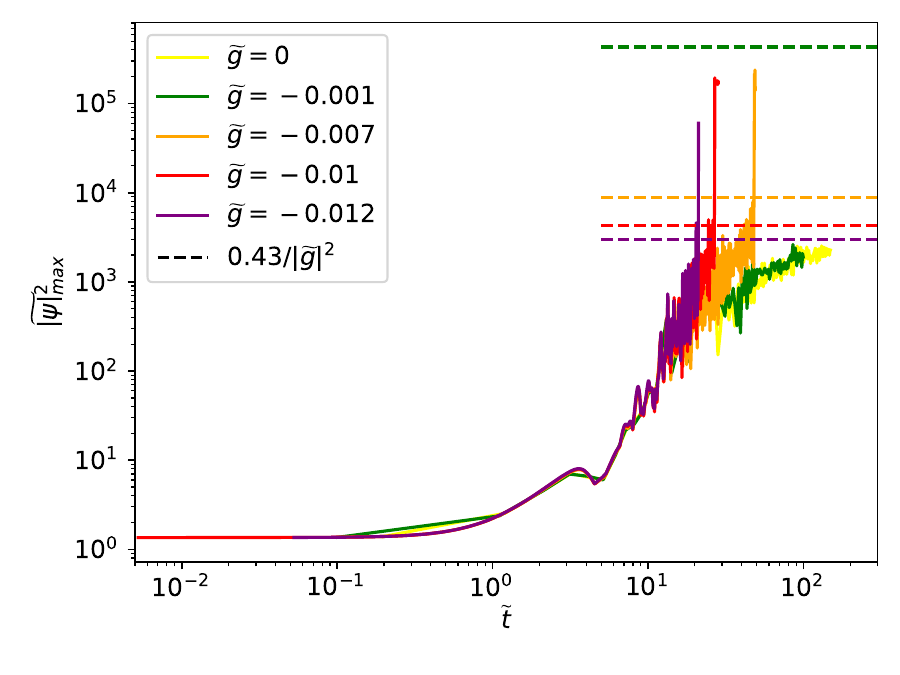}
\caption{Maximum density growth with respect to time from simulations assuming different self-interaction couplings: $\widetilde{g} = 0$ (without self-interactions), $\widetilde{g} = -0.001$, $\widetilde{g} = -0.007$, $\widetilde{g} = -0.01$, and $\widetilde{g} = -0.012$. The box size $\widetilde{L}=18$ and the total mass $\widetilde{N}=1005.3$.}
\label{fig:maxdenisty_small_g}
\end{figure}

The evolution of maximum density from simulations with different strength of self-interactions compared with the case without self-interactions is shown in Fig.~\ref{fig:maxdenisty_small_g}. These results support the theoretical prediction of~\cite{Levkov:2018kau} that gravity dominates the system and the effect of self-interactions is negligible in the early stages of boson star evolution.

As the central density continues to grow, however, the effect of self-interactions becomes increasingly important. We thus find that at large values of $|\widetilde{g}|$, the boson stars collapse at a critical mass, see Fig.~\ref{fig:maxdenisty_small_g} \cite{2011PhRvD..84d3531C,Chavanis:2011zm,Chavanis:2016dab,Levkov_2017}. Above the critical mass, the boson star is unstable to perturbations. The attractive self-interaction in Eq.~\ref{eq:GPP1} overcomes the quantum pressure, and boson stars shrink at an accelerated pace, developing huge boson densities in the center when maximum density reach the critical value, $\sim 0.43/\widetilde{g}^2$. Combining the relationship of Eq.~\ref{eq:scaling}, we know the critical mass of collapse is inversely proportional to $\sqrt{\widetilde{g}}$, in accordance with the theoretical critical mass, $M_{cr} \propto 1/(m\sqrt{g})$~\cite{Chavanis:2011zm,Chavanis:2016dab} (see also in \cref{APPE:soliton}).

\subsection{Bosons with repulsive self-interactions}
\label{boson_Repulsive_Self_Inter}
In this subsection, we study the evolution of some other candidates for dark matter, bosons with repulsive self-interactions \footnote{The linear theory of bosons with repulsive self-interactions, and constraints on the allowed interaction strength of DM, are studied in Ref.~\cite{Cembranos_2018}.}. 

\begin{figure}[htbp]
\centering
\textbf{Solitons with repulsive self-interactions}\par\medskip
\includegraphics[width=\columnwidth]{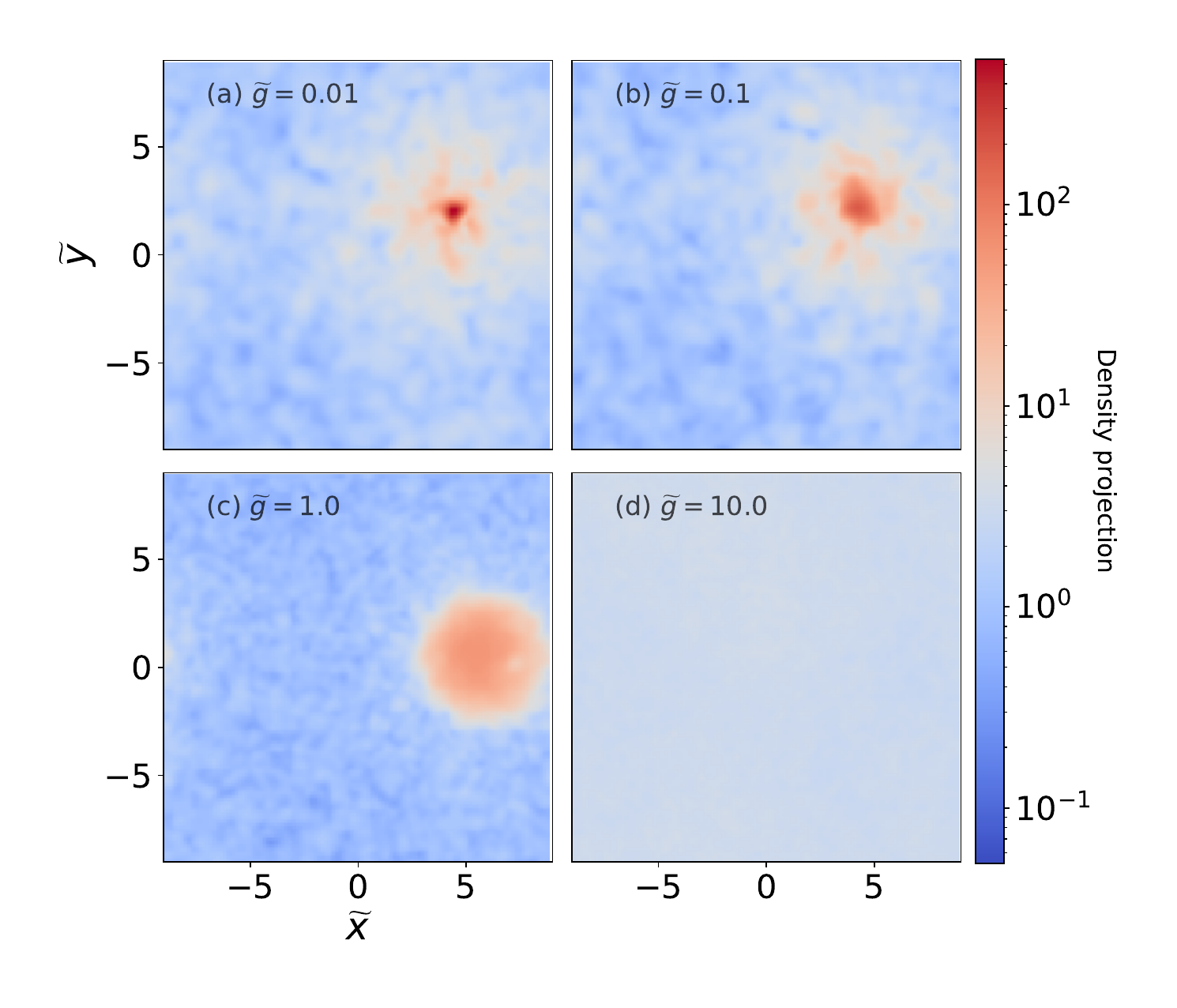}
\caption{Snapshots of the density field at $\widetilde{t} = 200$ from simulations in a box of size $\widetilde{L}=18$ and total mass $\widetilde{N}=1005.3$ assuming different $\widetilde{g}$.}
\label{fig:densityprojection_positive_g}
\end{figure}

By simulating the GPP equations with different positive values of $\widetilde{g}$ in a box of size $\widetilde{L}=18$ and total mass $\widetilde{N}=1005.3$, we find miniclusters form and dense objects appear in the center of the miniclusters for sufficiently weak $\tilde{g}$, see Fig.~\ref{fig:densityprojection_positive_g}(a-c).

\begin{figure}[htbp]
\centering
\includegraphics[width=\columnwidth]{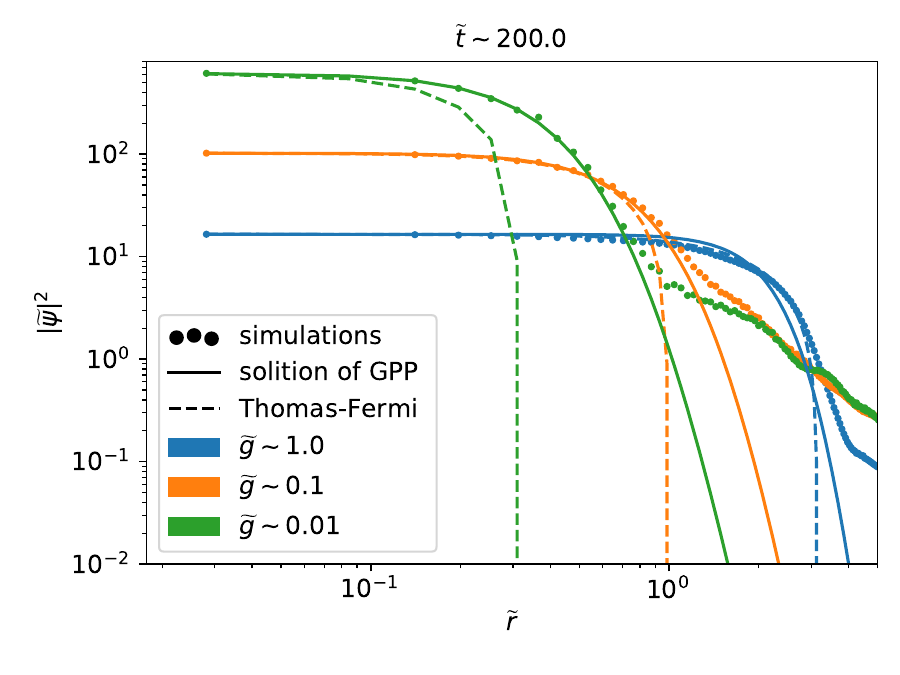}
\caption{Density profiles of the miniclusters from simulations (colored dots), compared with solitonic profiles (solid lines) as given by Eq.~\eqref{eq:soliton_profile} and Thomas-Fermi approximation ~\cite{Maga_a_2012} [Eq. (\ref{eq:rho_thomas_fermi})] with the same central densities.}
\label{fig:densityprofile_positiveg}
\end{figure}

The density profiles of the dense objects in the cases with $\widetilde{g} = 0.01$ and $\widetilde{g} = 0.1$ are shown in Fig.~\ref{fig:densityprofile_positiveg}, which can be well fit by the density profiles of solitons given by Eq.~\eqref{eq:sol_rho_gen}. Thus, we 
confirm solitons are condensed in the minicluster. 
We also prove the kinetic-energy term in Eq.~\ref{GPP_Equations} can be neglected when $\widetilde{g} \gtrsim 1.0$ since the density profile of the boson star becomes tantamount to the Thomas-Fermi approximation, Fig.~\ref{fig:densityprofile_positiveg}. Furthermore, as can be seen in Fig.~\ref{fig:densityprojection_positive_g} (d), we find that for very large repulsive self-interaction, $\widetilde{g}=10$, no boson star forms at all. In this case, the self-interaction dominates over gravity. Due to limited box size, the system forms a uniform condensate instead of a boson star.

\begin{figure}[htbp]
\centering
\includegraphics[width=\columnwidth]{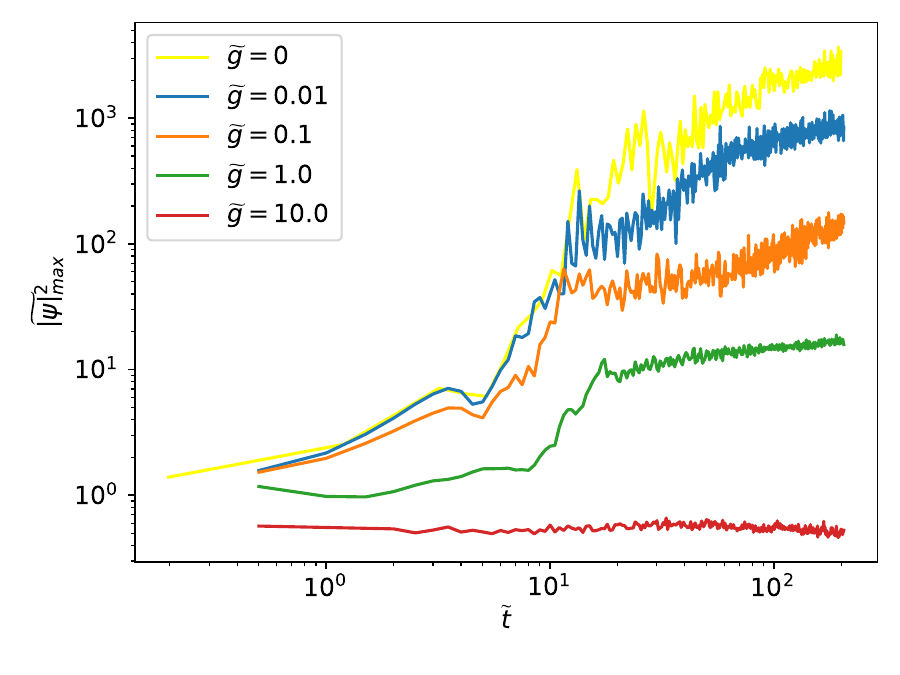}
\caption{Maximum density growth with respect to time from simulations assuming different $\widetilde{g}$. The box size $\widetilde{L}=18$ and the total mass $\widetilde{N}=1005.3$.}
\label{fig:maxdensity_positve_g}
\end{figure}

The evolution of maximum density with repulsive self-interactions is shown in Fig.~\ref{fig:maxdensity_positve_g}. The evolution of the maximum density with $\widetilde{g} = 0.01$ coincides with the case without self-interaction ($\widetilde{g} = 0$) at the early stage. This is similar to the case with weak attractive self-interactions. But at later stages when $\widetilde{t}>\widetilde{\tau}_{\rm sat}$, the growth rate of the maximum density is different.
The growth rate decreases with increasing $\widetilde{g}$ as expected.

\begin{figure}[htbp]
\centering
\includegraphics[width=\columnwidth]{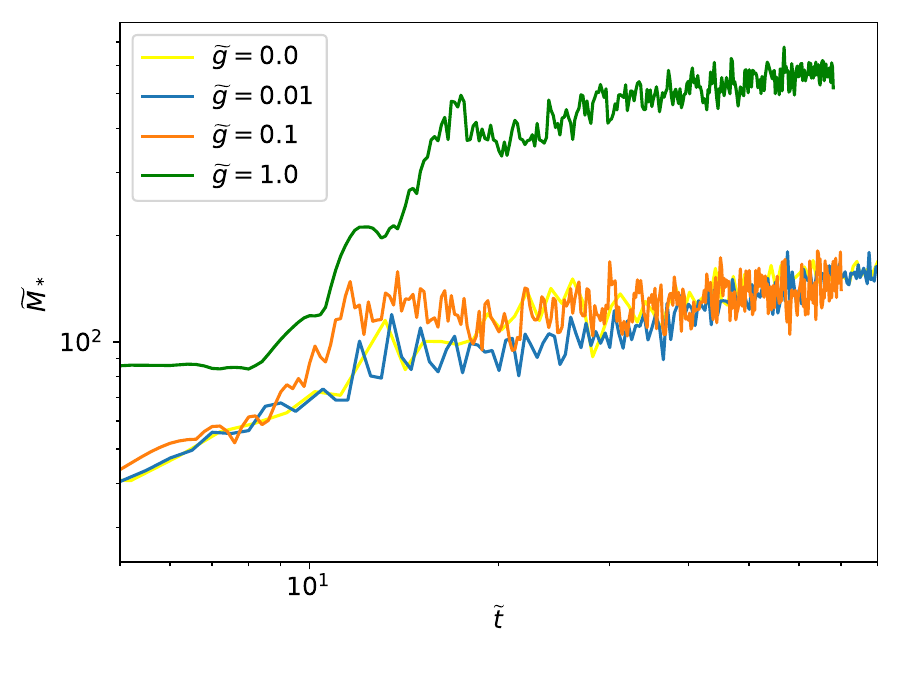}
\caption{Mass growth of boson stars with respect to time from simulations assuming different $\widetilde{g}$. The box size $\widetilde{L}=18$ and the total mass $\widetilde{N}=1005.3$. Note that the case with $g=10$ is not shown because no boson star is formed in the box.}
\label{fig:starmss_positve_g}
\end{figure}

However, with repulsive self-interactions, the radius of the boson star is larger compared to the case with no self-interactions (see Fig. ~\ref{fig:densityprofile_positiveg}). Thus for boson stars with the same central density, the mass of the ones with repulsive self-interactions is larger. To quantify how many particles condense in different cases, we look at the mass growth of boson stars, Fig.~\ref{fig:starmss_positve_g}. We find that while the central density growth is slower for larger positive $g$ as shown in Fig.~\ref{fig:maxdensity_positve_g}, the mass growth of boson stars is actually faster with increasing $g$ indicating that repulsive self-interactions promote the condensation process.

\section{Formation of multiple boson stars}
\label{Powerful_Attractive_Self_Inter}
It is possible that bosons can have even larger values of attractive self-coupling. Thus studying the evolution of these bosons are necessary as well. For bosons with attractive self-interactions, we have shown in \cref{Self_Inter_boson_Star} that when $|\widetilde{g}|$ is very small, gravity dominates the early stage evolution in systems, and leads to the formation of a single boson star per box~\footnote{In cosmological simulations~\cite{Schive:2014hza,Veltmaat:2018dfz,Bar:2018acw}, one boson star forms in each halo as it separates out from the cosmic expansion during gravitational collapse. We have verified that this occurs also in our simulations with an expanding background spacetime.}. The situation can be very different if $|g|$ is very large, and self-interactions dominate the early stages of evolution~\cite{Amin_2019}. In order to analyze these systems, we first introduce the governing equation for linear overdensity $\delta\equiv\delta\rho/\rho$, where $\rho$ is the mean density. In Fourier space, the linear overdensity $\delta_k$ satisfies \cite{2011PhRvD..84d3531C,Chavanis:2011uv,Desjacques_2018}
\begin{equation}
    \ddot{\delta}_k = -\left(\frac{k^4}{4m^2}+\frac{g \rho k^2}{m^2} -4 \pi G \rho\right)\delta_k.
\end{equation}
Here we have neglected the Hubble friction term and assumed the cosmic scale factor varies slowly on time scales we are concerned with so that it can be treated as a constant.
It is easy to find that $\delta_k$ will grow exponentially when
\begin{equation}
    k_J^2 < 2 \rho g + 2 \sqrt{\rho (\rho g^2 + 4 \pi G m^2)},
\end{equation}
i.e. the growth of the linear perturbation is unstable, thus the overdense regions will quickly undergo nonlinear collapse.

The instability scale $k_J$ is determined by the strength of gravity and self-interactions.
For different values of $g$ and $\rho$, we have:
\begin{itemize}
    \item $\rho g^2 \ll 4 m^2 \pi G$. Gravity dominates, miniclusters form first. After that, one boson star forms in the center of each minicluster.
    \item $\rho g^2 \approx 4 m^2 \pi G$. Gravity and self-interactions both play important roles. A gravitational bound minicluster may contain multiple boson stars formed from local overdensities. 
    \item $\rho g^2 \gg 4 m^2 \pi G$. Self-interactions dominate. The condensate can fragment and form multiple boson stars due to self-interactions before a gravitational bound object forms.
\end{itemize}

To test this hypothesis, we run simulations with very strong attractive self-couplings.
For comparison, we also simulate the Gross-Pitaevskii (GP) equations ignoring gravity:
\begin{eqnarray}
i\frac{\partial}{\partial{\widetilde{t}}}\widetilde{\psi}&=&-\frac{1}{2}\widetilde{\nabla}^2\widetilde{\psi} +\widetilde{g}|\widetilde{\psi}|^2\widetilde{\psi}
\label{eq:GP}
\end{eqnarray}
under the same initial conditions.

\begin{figure}[htbp]
\centering
\textbf{Condensate fragmentation at large attractive coupling}\par\medskip
\includegraphics[width=\columnwidth]{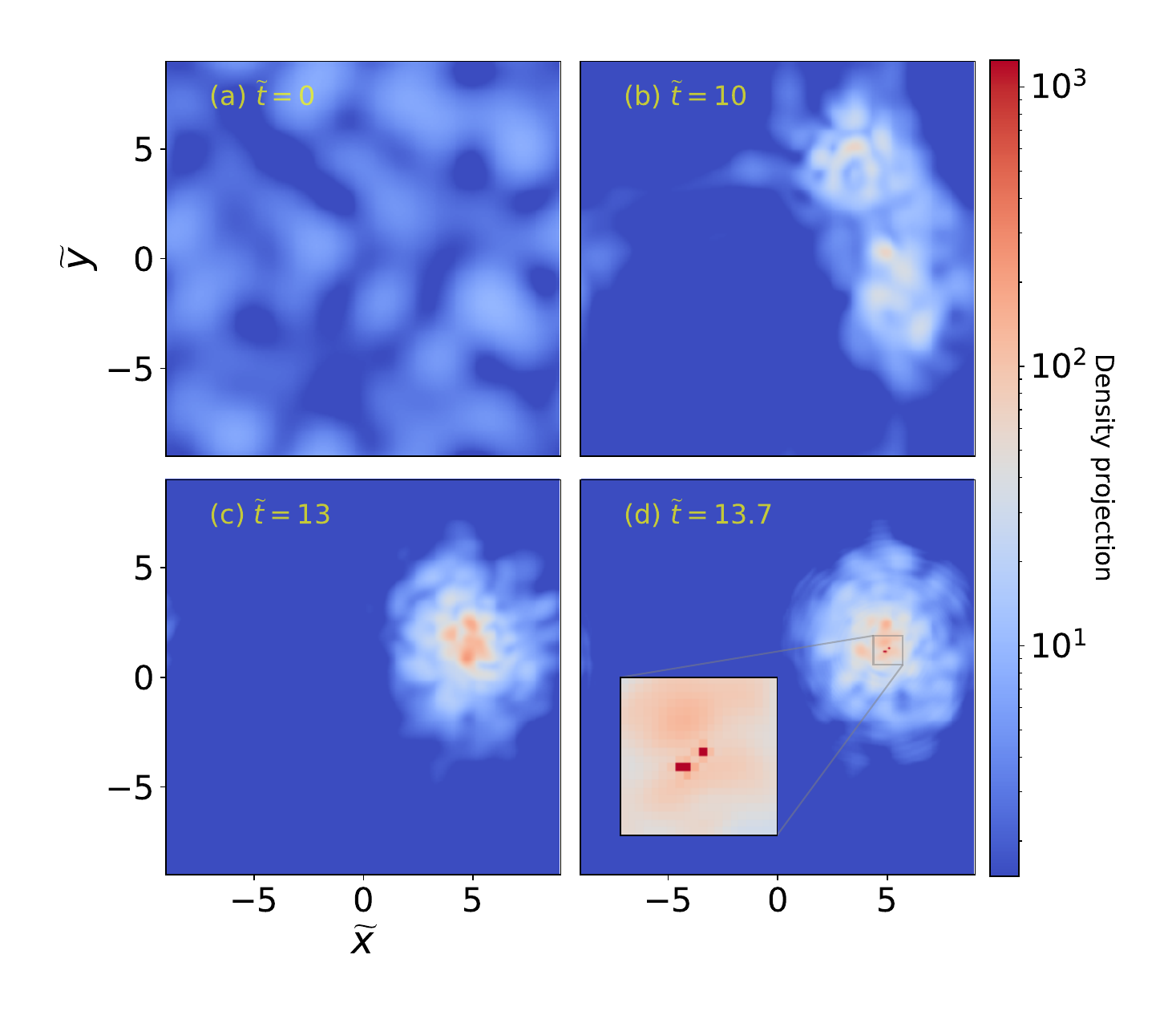}
\caption{Snapshots of the density field from simulations of the GPP equations in a box of size $\widetilde{L}=18$, and total mass $\widetilde{N} = 1005.3$. The self-interaction coupling constant $\widetilde{g} = -0.04$. Note that due to resolution limit, we can not resolve the central region of the densest object, so we cutoff the projected density in the plot at $\approx 1500$.}
\label{fig:densityprojection_g-0point04}
\end{figure}

Fig.~\ref{fig:densityprojection_g-0point04} shows the evolution of the system simulated using GPP equations with $\widetilde{g} = -0.04$. We can see the formation of a minicluster, Fig.~\ref{fig:densityprojection_g-0point04} (a-c). After that, several boson stars form in the system, Fig.~\ref{fig:densityprojection_g-0point04} (d).

\begin{figure}[htbp]
\centering
\includegraphics[width=\columnwidth]{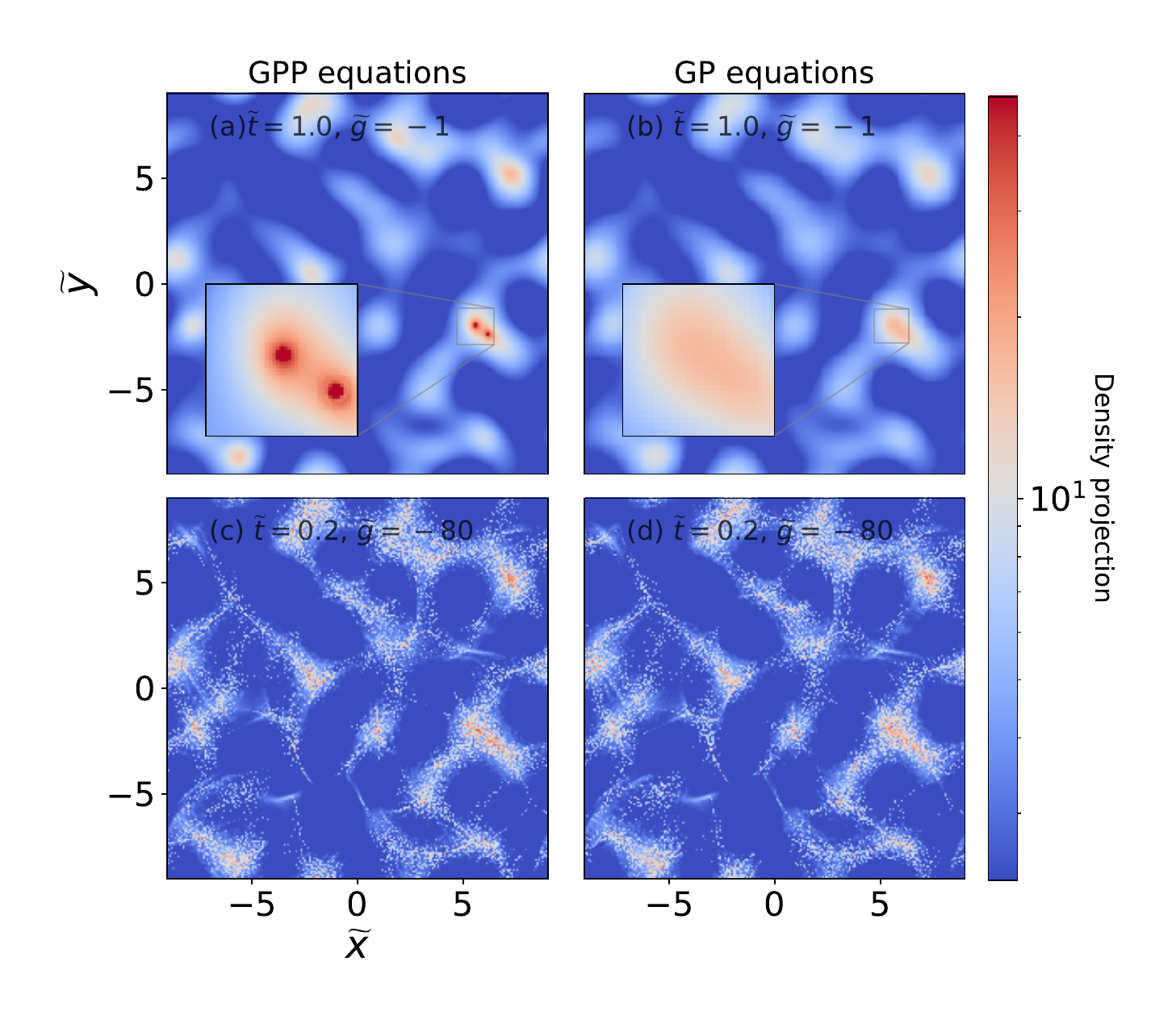}
\caption{Snapshots of the density field from simulations of the GPP equations (left column) and GP equations (right column). We pick $\widetilde{g}=-1$ (first row) and  $\widetilde{g}=-80$ (second row). The box size $\widetilde{L}=18$, and the total mass $\widetilde{N}=1005.3$.}
\label{fig:densityprojection_with_without_gravity}
\end{figure}

Fig.~\ref{fig:densityprojection_with_without_gravity} (a) and (b) show the systems simulated using GPP equations and GP equations with $\widetilde{g} = -1.0$ at $\widetilde{t} = 1.0$. We can see two boson stars condense in the dense areas in Fig.~\ref{fig:densityprojection_with_without_gravity} (a), but not in (b), suggesting that the gravity can promote the condensation of boson stars slightly even when self-interactions are strong.

Fig.~\ref{fig:densityprojection_with_without_gravity}  (c) and (d) show the cases with $\widetilde{g} = -80$. Comparing results from GPP equations with the ones from the GP equations, we don't find a big difference. Therefore, we conclude that the self-interactions dominate the evolution of boson stars alone in some extreme systems.

In fact, Eq.~\eqref{eq:time_compare} shows the self-interactions can be ignored if $-0.53 \lesssim \widetilde{g}<0$ for a system with box size $\widetilde{L}=18$, total mass $\widetilde{N}=1005.3$, and characteristic velocity $\widetilde{v} \sim 1$. But our simulation shows self-interactions are important even for $\widetilde{g} = -0.04$ at the late stages of evolution (see Fig. \ref{fig:densityprojection_g-0point04} (c) and (d)). We think the reason is that at these times, the characteristic velocity increases due to gravitational collapse making the gravitational condensation less efficient.

In a cosmological setting, the extreme condensate fragmentation observed in our simulations caused by strong self-interactions would spoil the hierarchical nature of cosmic structure formation. However, these results could be applicable to fragmentation of the inflation condensate (e.g. Ref.~\cite{Niemeyer_2020}) or to condensates in condensed matter.

\section{Conclusions}
\label{Conclusion}
 
By means of numerical solution of the dynamical Gross-Pitaevskii-Poisson equations, we studied the formation and subsequent growth of boson stars inside gravitationally self-bound halos. We demonstrated a series of new phenomena in the solutions, which had not been seen before in the dynamical regime.

In the case with no self-interactions beyond gravity, we demonstrated the saturation of boson star growth. We ran simulations for times long compared to the dynamical timescales, i.e. $t\gg\tau_{\rm sat}\gg \tau_{\rm gravity}$, and much longer than those of Ref.~\cite{Levkov:2018kau}. In this regime of boson stars we observed a transition from relatively fast mass growth, $\propto t^{1/2}$, to much slower growth, $\propto t^{1/8}$, in accordance with the prediction made by Ref.~\cite{Eggemeier:2019jsu}. We attribute this to the formation of a gravitationally bound and virialised atmosphere around the boson star, suppressing further mass growth by coupling the condensation time to the boson star's virial temperature.

Another interesting phenomenon is that we discover no significant difference for the end stage evolution of maximum density (see Fig.~\ref{fig:1005_18_max_density_tog}) normalized by $\widetilde{N}^2$, $\widetilde{N}^{4/3}$, $\widetilde{N}^{1.5}$, with all values in this range providing similarly good fits to the data. Although, as mentioned, the $\widetilde{N}^{4/3}$ normalization could be explained by the core-halo mass relation, $M_c\propto M_h^{1/3}$~\cite{Schive:2014hza}, we cannot be certain that this relation with the exponent $1/3$ holds in our simulations. Thus, this is an interesting topic for future research.

In any case, our observation of a reduced boson star growth rate at late times explains why boson stars in virialised halos in cosmological simulations (e.g. Refs.~\cite{Schive:2014dra,Veltmaat:2018dfz,Eggemeier:2019jsu}) are only observed to grow very slowly compared to the other gravitational timescales, and thus populate an almost constant in time core-halo mass relation (see also Ref.~\cite{Du:2016aik}, which considers the effect of mergers). 

Our results in the case of attractive self-interactions demonstrated for the first time that boson stars can grow via accretion and reach the critical mass for collapse. Once the critical mass is reached, relativistic simulations are needed. The relativistic simulations of Refs.~\cite{2017PhRvL.118a1301L,2017JCAP...03..055H} began with super-critical stars, and showed that these stars lead to either ejection of relativistic bosons and a massive remnant (nova), or, for weak self-interactions, collapse to black holes. Our dynamical simulations show that it is possible to reach such critical nova state dynamically before saturation. This implies that such a star could undergo a series of novae in its lifetime. This could have implications for the abundance of relativistic particles in the Universe. If the bosons produced can be converted into visible photons, as is the case for axions and axion-like particles, the nova ejecta could even be observed. We leave for future work the study of he expected rates in realistic models.
 
In the case of very strong attractive interactions we demonstrated that these can dominate over gravity and lead to fragmentation of the condensate into many small, dense regions. Such fragmentation has not been seen before in simulations including gravity. This has implications for the fragmentation of the inflation condensate during the reheating epoch~\cite{Musoke:2019ima,Niemeyer_2020}.
 
Our results in the case of repulsive interaction demonstrated that such an interaction can promote boson star formation. We showed for the first time that the stable Thomas-Fermi-like solution, which has been studied often in the literature on scalar field DM (e.g. Refs.~\cite{Maga_a_2012}), can be reached dynamically via gravitational accretion. Repulsive self-interactions change the mass-radius relation of boson stars, and we have shown that these solitons can also be formed dynamically via condensation. A realisitic formation mechanism for such states also has implications for the gravitational wave searches for exotic compact objects~\cite{Giudice:2016zpa}, and could be used to predict the expected signal rates in gravitational wave detectors \cite{Clough:2018exo}. 

In summary, we have demonstrated new results on the dynamical formation and growth of boson stars in a collection of different models, including self-gravity, attractive and repulsive self-interactions. Our results have applications to future terrestrial, astrophysical, and cosmological observations searching for new types of bosons across a wide range of scales.

\section{Acknowledgements}
\label{Acknow}
We thank B. Schwabe, J. H. H. Chan, M. Gosenca, S. Hotchkiss, and R. Easther for helpful discussions. We would also like to thank L. Visinelli, Z. Liu, Z. Chen, D. Chen, Y. Huo, X. Kong, H. Sebastian for constructive comments which helped to improve this paper. X. Du thanks Chanda Prescod-Weinstein for beneficial discussions. DJEM thanks Luca Visinelli for useful discussions. J. Chen acknowledges the China Scholarship Council (CSC) for financial support. X. Du acknowledges support from NASA ATP grant 17-ATP17-0120. DJEM and EWL are supported by the Alexander von Humboldt Foundation and the German Federal Ministry of Education and Research.

\appendix

\section{Pseudo-spectral Method}
\label{APPE:Fourth_Order_PS}
To solve the SP, GP, and GPP equations, we use a fourth-order time-splitting pseudospectral solver with GPU acceleration~\cite{cuda_gpu}. Compared to the four-order pseudospectral method used in~\cite{Du:2018qor}, our code is $6-7$ times faster under the same resolution. 

The wave function is advanced in time by a unitary operator,
\begin{equation}
\psi(t+\Delta t)=e^{-i\hat{H} \Delta t}\psi=e^{-i(K+W) \Delta t}\psi,
\label{eq:psi_evolve}
\end{equation}
where the Hamiltonian operator is split into the kinetic part $K$ and the potential part $W$. $\Delta t$ is the time step size. 
In general, the operator $e^{-i(K+W)\Delta t}$ can be expanded as
\begin{equation}
e^{-i(K+W)\Delta t}=\prod_{j}^N({e^{-ik_j \Delta t {K}}}{e^{-iv_j \Delta t {V}}}),
\end{equation}
where $k_j$ and $v_j$ are constant parameters which are determined by requiring that the expansion is accurate up to a specified
order. For example, to the second order, we obtain the well-known leapfrog method,
\begin{eqnarray}
\!\!\!\!\!\!e^{-i(K+W)\Delta t} = && e^{-i\frac{1}{2}W \Delta t} e^{-iK \Delta t} e^{-i \frac{1}{2}W \Delta t}+\mathcal{O}(\Delta t^3),
\end{eqnarray}
which is also referred to as the ``kick-drift-kick” scheme. In our simulations, we implement the fourth-order algorithm proposed by~\cite{McLachlan:1995,chin2007forward}
\begin{eqnarray}
e^{-i(K+W)dt} = && ...e^{-i v_0 \Delta t W} e^{-i t_1 \Delta t K}e^{-i v_1 \Delta t W} \nonumber \\
                &&    e^{-i t_2 \Delta t K} e^{-i v_2 \Delta t W} + \mathcal{O}(\Delta t^5),
\label{M1_forth}
\end{eqnarray}
where $v_0$, $t_1$, $v_1$, $t_2$, $v_2$ are parameters, and $...$ means operations symmetric with the right terms in the equation. These parameters are given by:
\begin{eqnarray}
w  &=& \sqrt{3-12v_1+9v_1^2},\\
t_1 &=& \frac{1}{2}-t_2,\\
t_2 &=& \frac{1}{4}\left(1-\sqrt{\frac{9v_1-4+2w}{3v_1}}\right),\\
v_0 &=& 1-2(v_1+v_2),\\
v_1 &=& \frac{121}{3924}(12-\sqrt{471}),\\
v_2 &=& \frac{1}{6}-4v_1 t_1^2.
\end{eqnarray}
For the time step size, $\Delta t$, we require~\cite{Woo:2008nn}:
\begin{equation}
\Delta t= {\rm min}\left\{{\frac{m \Delta x^2}{6 \pi},\frac{\pi}{4 m V_{\rm max}}}\right\},
\label{eq:deltat}
\end{equation}
where $V_{\rm max}$ is the maximum absolute value of the potential and $\Delta x$ is spatial cell size.

\section{Convergence analysis}
\label{APPE:Conv}
\subsection{Temporal resolution}
\label{APPE:Time_Res_Conv}
\begin{figure}[htbp]
\centering
\includegraphics[width=\columnwidth]{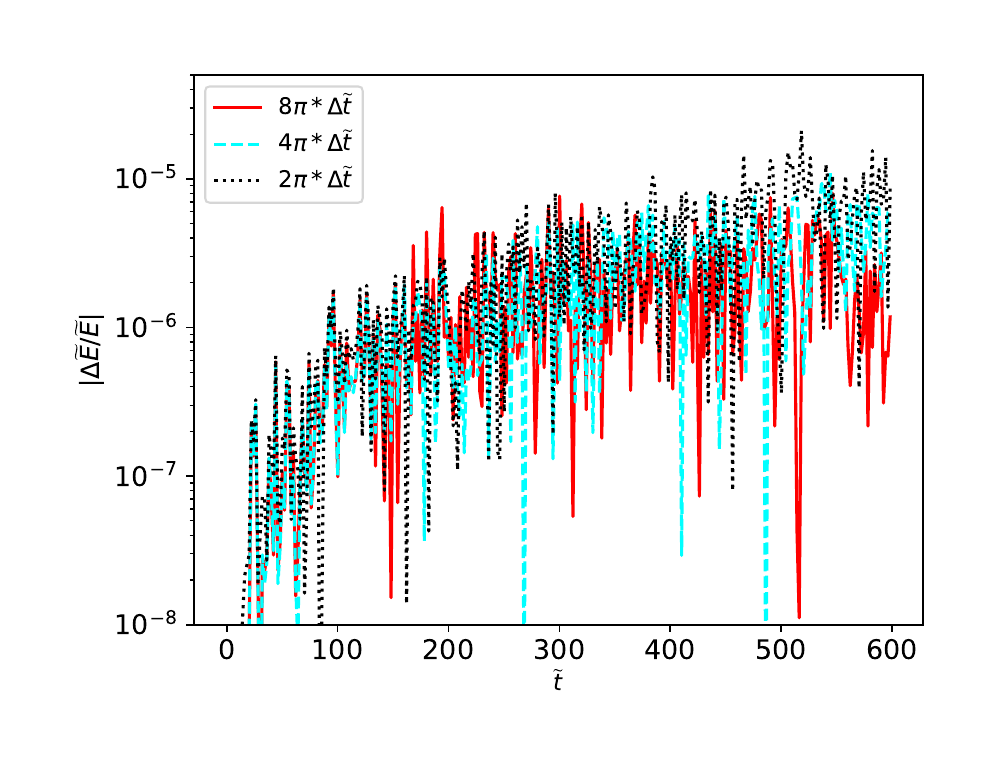}
\caption{Relative errors of the total energy with respect to time for different time step sizes. To show the fourth-order convergence of the algorithm in time, we have multiplied the relative errors by a factor of $2^4$ and $4^4$ for the cases
with time step sizes $4\pi\Delta \widetilde{t}$ and $ 2\pi\Delta \widetilde{t}$, respectively.}
\label{fig:time_resolution}
\end{figure}

To test the convergence of our code in time, we run one typical simulation with decreasing time step sizes and examine the conservation of total energy. The box has a length of $18$ on each side and is simulated with $256^3$ cells. The bosons in the box have only gravitational interaction, i.e. $\widetilde{g} = 0$, and have a total mass of $1005.3$. We find that the total energy is fourth-order conserved as expected (see Fig.~\ref{fig:time_resolution}).

\subsection{Spatial resolution}
\label{APPE:Space_Res_Conv}
\begin{figure}[htbp]
\centering
\includegraphics[width=\columnwidth]{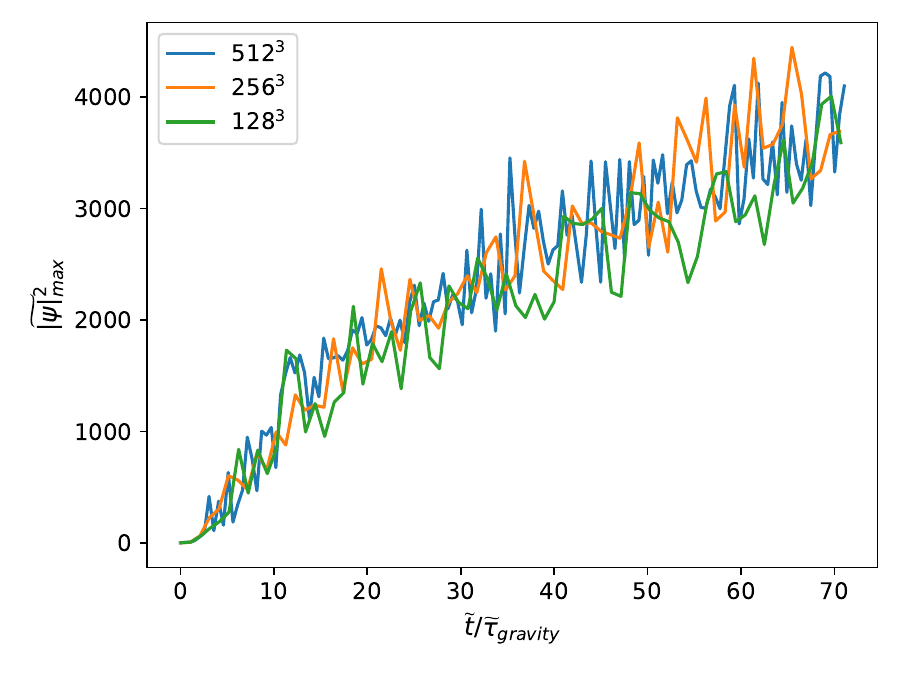}
\caption{Maximum density growth with respect to time for spatial resolutions.}
\label{fig:space_resolution}
\end{figure}

We run the same simulation in Sec.~\ref{APPE:Time_Res_Conv} with different spatial resolutions: $128^3$, $256^3$, and $512^3$.
The maximum density in the box as a function of time is shown in Fig.~\ref{fig:space_resolution}. As can be seen, the results from
the lowest-resolution run is consistent with the highest-resolution run, suggesting that even with a resolution as low as $128^3$. we
can already get reliable results.

\begin{figure}[htbp]
\centering
\includegraphics[width=\columnwidth]{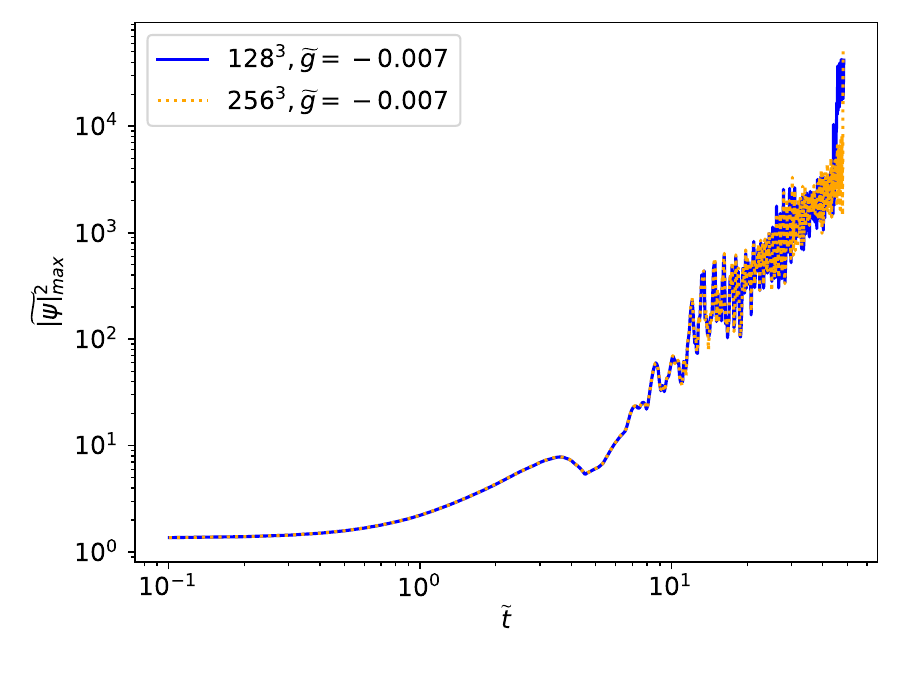}
\caption{Maximum density with respect to time for different spatial resolutions. The bosons have
attractive self-interaction, i.e. $\widetilde{g}= -0.007$.}
\label{fig:space_reso_gpp}
\end{figure}

We also check the cases with self-interactions. The initial conditions are taken to be the same, but the simulations are run with 
$\widetilde{g} = -0.007$. The maximum density is shown in Fig.~\ref{fig:space_reso_gpp}. Again, we can see that the results are converged as the resolution increases. At late times when the central density of the boson star approaches to the critical value, the maximum density has a rapid increase. This happens slightly later in the high-resolution run suggesting that we may not have enough resolution at that time. But in this paper, we will focus on the growth of boson star before the critical collapse.

\section{Soliton solutions to the GPP equations}
\label{APPE:soliton}

Plugging the ansatz of stationary solution
\begin{equation}
\psi(\mathbf{r},t) =\psi(\mathbf{r})e^{-i E t}
\label{eq:ansatz}
\end{equation}
into Eqs. (\ref{eq:GPP1}) and (\ref{eq:GPP2}), we get the time-independent GPP equations

\begin{eqnarray}
-\frac{1}{2}\nabla^2\psi(\mathbf{r})&=&\left[E-V(\mathbf{r})-g|\psi(\mathbf{r})|^2\right]\psi(\mathbf{r}),
\label{eq:GPP_ind_1}
\\
\nabla^2V(\mathbf{r})&=&|\psi(\mathbf{r})|^2.
\label{eq:GPP_ind_2}
\end{eqnarray}
Here we have written the equations in dimensionless form as in Sec. \ref{GPP_Equations} and dropped the tildes over the
dimensionless quantities for simplicity. The soliton solution is the eigenstate of Eqs. (\ref{eq:GPP_ind_1}) and
(\ref{eq:GPP_ind_2}) with the lowest eigenenergy under the boundary conditions
\begin{eqnarray}
\psi(0)&=&\psi_0,\label{eq:bnds1}\\
\psi'(0)&=&0,\label{eq:bnds2}\\
\psi(\infty)&=&0,\label{eq:bnds3}\\
V'(0)&=&0,\label{eq:bnds4}\\
V(\infty)&=&0.\label{eq:bnds5}
\end{eqnarray}
In practice, Eqs. (\ref{eq:GPP_ind_1}) and (\ref{eq:GPP_ind_2}) can be solved numerically using the shooting method:
(1) let $V(0)=V_0$, $E=E_0$ and integrate the equations outward from $r=0$; (2) adjust the values of $V_0$
and $E_0$ until the boundary conditions Eqs. (\ref{eq:bnds3}) and (\ref{eq:bnds5}) are satisfied.

Note that the GPP equations have the following scaling symmetry
\begin{equation}
\{\mathbf{r},t,\psi,E, V, g\} \rightarrow \{\lambda^{-1} \mathbf{r}, \lambda^{-2} t, \lambda^2\psi, \lambda^2 E, \lambda^2 V, \lambda^{-2} g\},
\label{eq:scaling}
\end{equation}
where $\lambda$ is an arbitrary none-zero parameter. Using this scaling symmetry, we can transform one soliton solution
to another solution that has a different central density $\rho_0=|\psi_0|^2$ but the same $g^2\rho_0$.

For a scalar field without self-interaction, i.e. $g=0$, it has been shown that the density profile of a soliton can be
well fit by \cite{Schive:2014hza,2015MNRAS.451.2479M}
\begin{equation}
\rho_{\rm soliton}(r)=\rho_0\left[1+0.091\left(\frac{r}{r_{\rm core}}\right)^2\right]^{-8}.
\label{eq:soliton_profile}
\end{equation}
Here only one of the parameters $\rho_0$ and $r_{\rm core}$ is independent. Given $\rho_0$, the core radius, defined as the radius where
the density drops to half of the central density,
$r_{\rm core}=1.308\rho_0^{-1/4}$. The soliton is uniquely determined by its central density.

When the self-interaction is non-negligible, we will need an additional parameter $g^2\rho_0$ to determine the soliton profile.

As $g^2\rho_0$ approaches $0$, we expect that the soliton has the same density profile as Eq. (\ref{eq:soliton_profile}).
So we assume in the general case the soliton density profile has a form of
\begin{equation}
\rho_{\rm soliton}(r)=\rho_0\left[1+(-1+2^{1/\beta})\left(\frac{r}{r_{\rm core}}\right)^{\alpha}\right]^{-\beta},
\label{eq:sol_rho_gen}
\end{equation}
where $\alpha$ and $\beta$ are functions of $g^2\rho_0$ only. When $g\rho_0^2\rightarrow 0$, we require that
$\alpha\rightarrow2$, and $\beta\rightarrow8$.

\begin{figure}[htbp]
\centering
\includegraphics[width=\columnwidth]{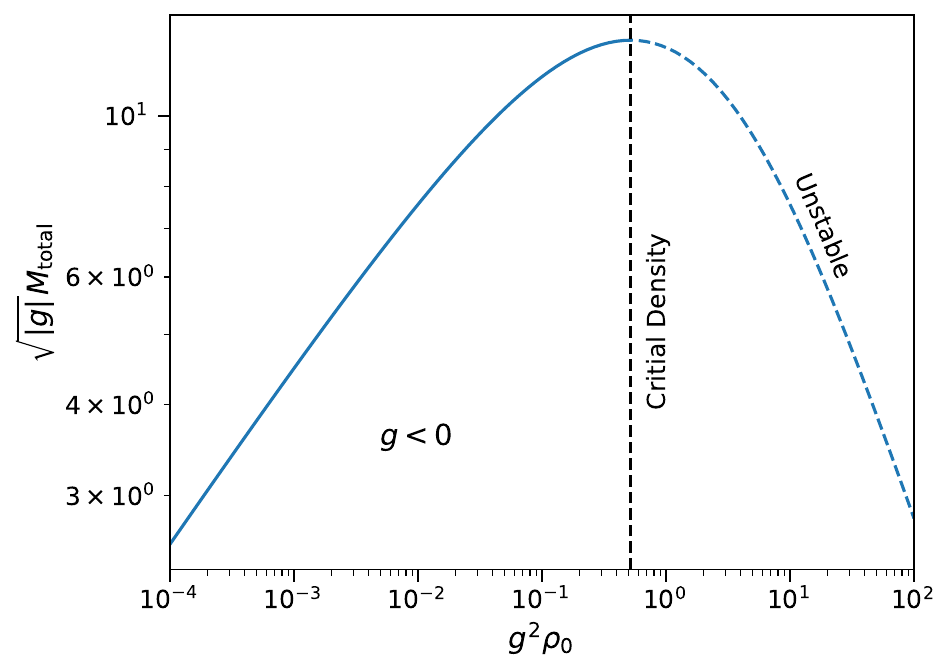}
\caption{Total mass of boson star with attractive self-interactions ($g<0$) as a function of the central density.}
\label{fig:M_rho_g_neg}
\end{figure}

We first consider the case with attractive self-interactions, i.e. $g < 0$. As is well known, there exists a critical
mass above which a boson star with attractive self-interactions is unstable ~\cite{2011PhRvD..84d3531C,Chavanis:2011zm}. In Fig.~\ref{fig:M_rho_g_neg},
we show the total mass of the boson star, $\sqrt{|g|} M_{\rm total}$, with respect to its central density, $g^2\rho_0$.
As expected, $\sqrt{|g|} M_{\rm total}$ increases with $g^2\rho_0$ and reached a maximum value, $12.72$, at $g^2\rho_0=0.52$.
When the central density increases further, the soliton solution becomes unstable and its total mass decreases as the
central density increases.

\begin{figure}[htbp]
\centering
\includegraphics[width=\columnwidth]{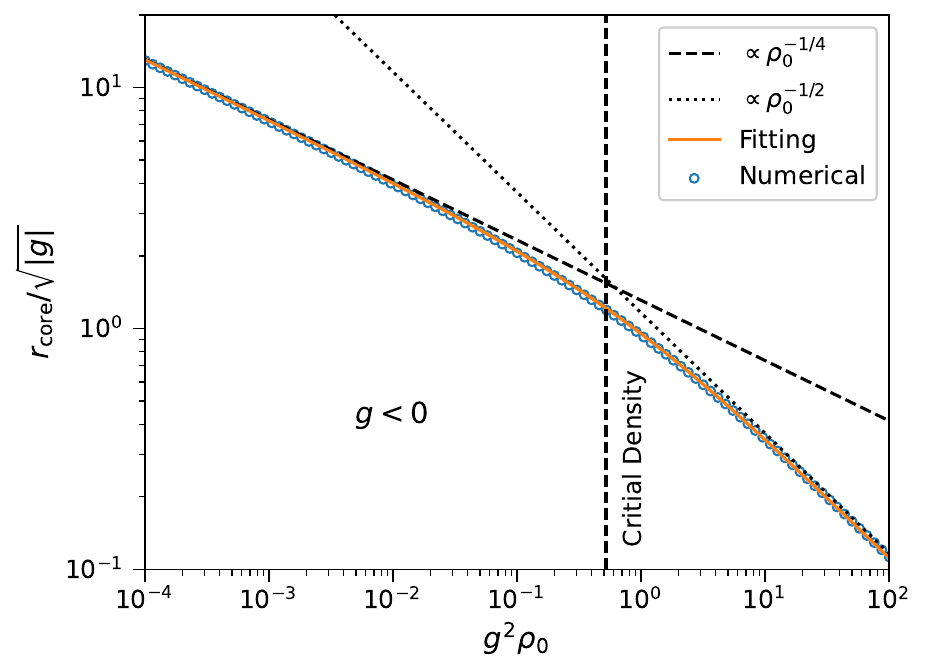}
\caption{Core radius of boson star with attractive self-interactions ($g<0$) as a function of the central density. Circles:
numeric results. Solid line: fitting function, Eq. (\ref{eq:r_core_g_neg}).}
\label{fig:r_c_rho_c_g_neg}
\end{figure}

To get the fitting formula for the density profile, we also need to know how the core radius depends on $g$ and $\rho_0$.
Figure~\ref{fig:r_c_rho_c_g_neg} shows the core radii of boson stars with different central densities. As can been seen, when $g^2\rho_0 \ll 1$,
the core radius $r_{\rm core}\propto \rho_0^{-1/4}$, recovering the relation seen in the case without self-interactions. When
$g^2\rho_0 \gg 1$, $r_{\rm core}\propto \rho_0^{-1/2}$. So we assume the core radius has the form
\begin{equation}
r_{\rm core} = 1.308\sqrt{|g|}\left[-\frac{a}{2}+\sqrt{\left(\frac{a}{2}\right)^2+\frac{1}{g^2\rho_0}}\right]^{1/2},
\label{eq:r_core_g_neg}
\end{equation}
where $a$ is a free parameter needed to be determined by fitting the numerical results. We find the best-fit value of
$a$ is $1.375$. Note that the solution with $g^2\rho > 0.52$ is unstable as discussed previously, but we include all
the solutions with $10^{-4}<g^2\rho_0<10^2$ in the fitting process so that we can correctly get the transition between
two limits: $r_{\rm core} \propto\rho_0^{-1/4}$ and $r_{\rm core} \propto\rho_0^{-1/2}$.

\begin{figure}[htbp]
\centering
\includegraphics[width=\columnwidth]{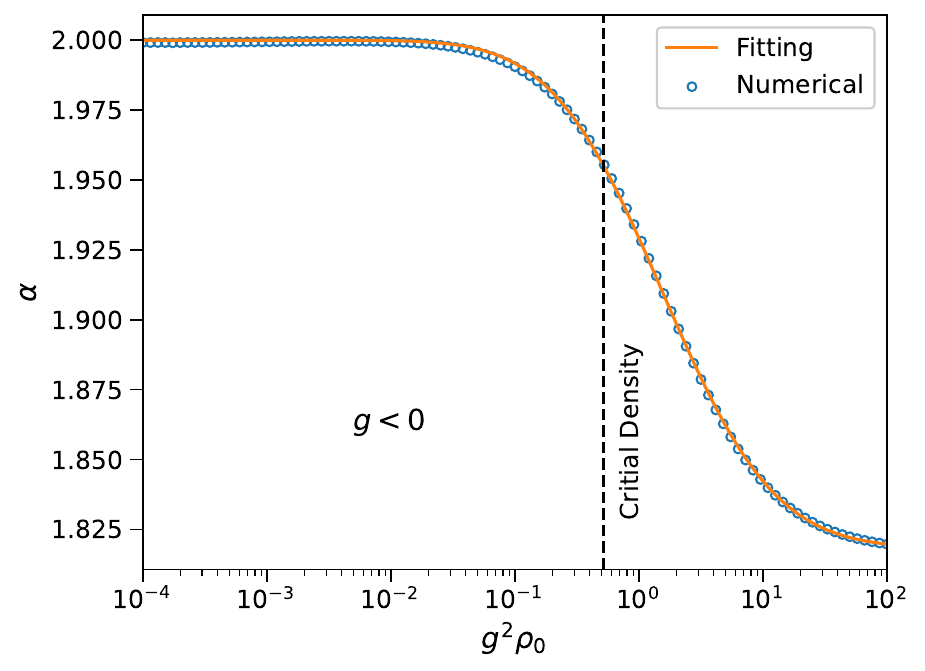}
\caption{Parameter $\alpha$ in the fitting formula Eq. (\ref{eq:sol_rho_gen}) for the case with attractive self-interactions ($g<0$). Circles: numeric results. Solid line: fitting function, Eq. (\ref{eq:alpha}).}
\label{fig:alpha_g_neg}
\end{figure}

\begin{figure}[htbp]
\centering
\includegraphics[width=\columnwidth]{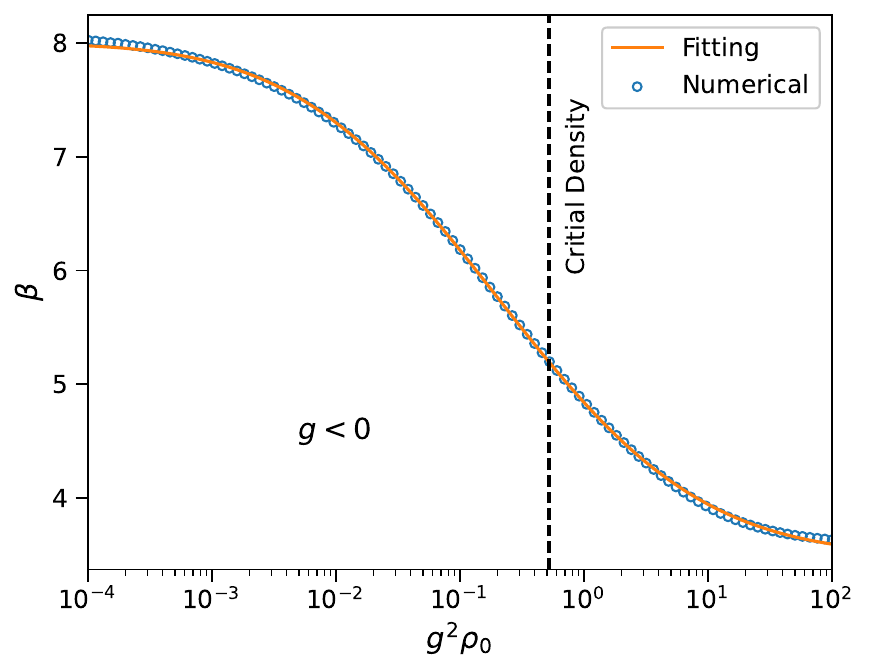}
\caption{Parameter $\beta$ in the fitting formula Eq. (\ref{eq:sol_rho_gen}) for the case with attractive interactions ($g<0$) as a function of the central density. Solid line: fitting function, Eq. (\ref{eq:beta}).}
\label{fig:beta_g_neg}
\end{figure}

To determine the parameter $\alpha$ and $\beta$ for each solution, we first fix the core radius using Eq.~(\ref{eq:r_core_g_neg}).
Then we fit $r^2\rho(r)$ within the range $0.01 r_{\rm core} < r < 5 r_{\rm core}$. The best-fit $\alpha$ and $\beta$ for different
$g^2\rho_0$ are shown in Figs. \ref{fig:alpha_g_neg} and \ref{fig:beta_g_neg}, respectively. We find that the dependence of $\alpha$ and $\beta$ on $g^2\rho_0$ can be well fit by
\begin{eqnarray}
\alpha&=&\alpha_a+(2-\alpha_a)\tanh^8{\left[\alpha_b(g^2\rho_0)^{-\alpha_c}\right]},
\label{eq:alpha}\\
\beta&=&\beta_a+(8-\beta_a)\tanh^8{\left[\beta_b(g^2\rho_0)^{-\beta_c}\right]}
\label{eq:beta}.
\end{eqnarray}
The best-fit values for $\alpha_i$ and $\beta_i$ ($i=a,b,c$) are listed in Table \ref{tab:rho_fit}.

\begin{figure}[htbp]
\centering
\includegraphics[width=\columnwidth]{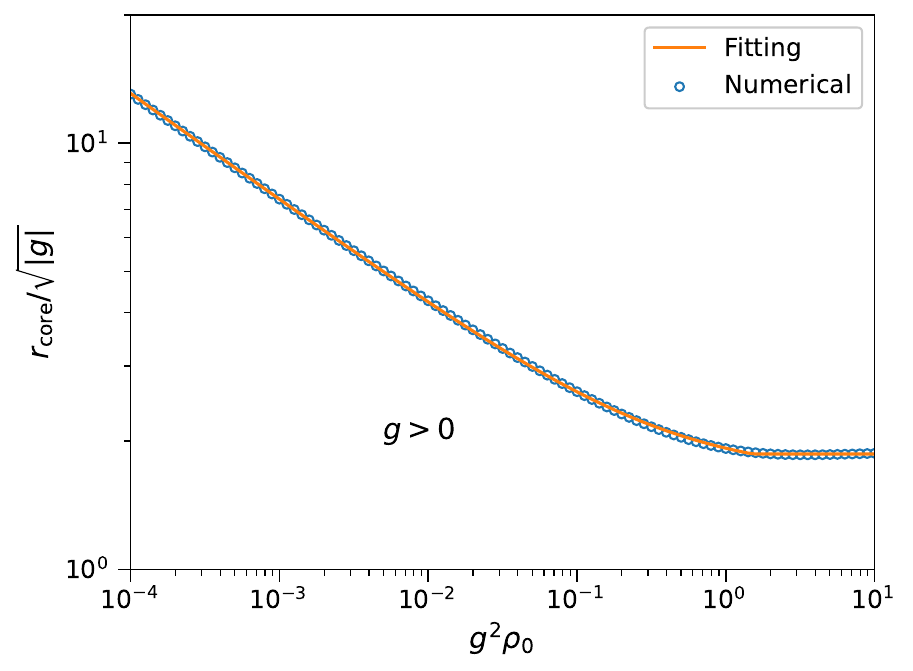}
\caption{Core radius of boson star with repulsive self-interactions ($g>0$) as a function of the central density.
Circles: numeric results. Solid line: fitting function, Eq. (\ref{eq:r_core_g_pos}).}
\label{fig:beta_g_pos}
\end{figure}

Similarly, we can find the relation between $r_{\rm core}$ and $\rho_0$ for the case with repulsive self-interactions ($g>0$).
We assume
\begin{equation}
r_{\rm core} = \left\{
\begin{aligned}
1.308\sqrt{|g|}\left[2+(g^2\rho_0)^{-b}\right]^{\frac{1}{4 b}}&,&g^2\rho_0 \leq 1.5, \\
c \sqrt{|g|} &,&g^2\rho_0 > 1.5,
\end{aligned}
\right.
\label{eq:r_core_g_pos}
\end{equation}
considering that when $g^2\rho\gg1$, we have Thomas-Fermi-like solution at small radii
\begin{equation}
\rho(r) = \rho_0\frac{\sin (r/\sqrt{g})}{r/\sqrt{g}},
\label{eq:rho_thomas_fermi}
\end{equation}
which gives a core radius that is independent of the central density. We find the best-fit $b=0.710752$, and $c=1.86543$.
Fig.~\ref{fig:beta_g_pos} shows the fitting formula of $r_{\rm core}$ (solid line) compared
with the numerical results (circles).

\begin{figure}[htbp]
\centering
\includegraphics[width=\columnwidth]{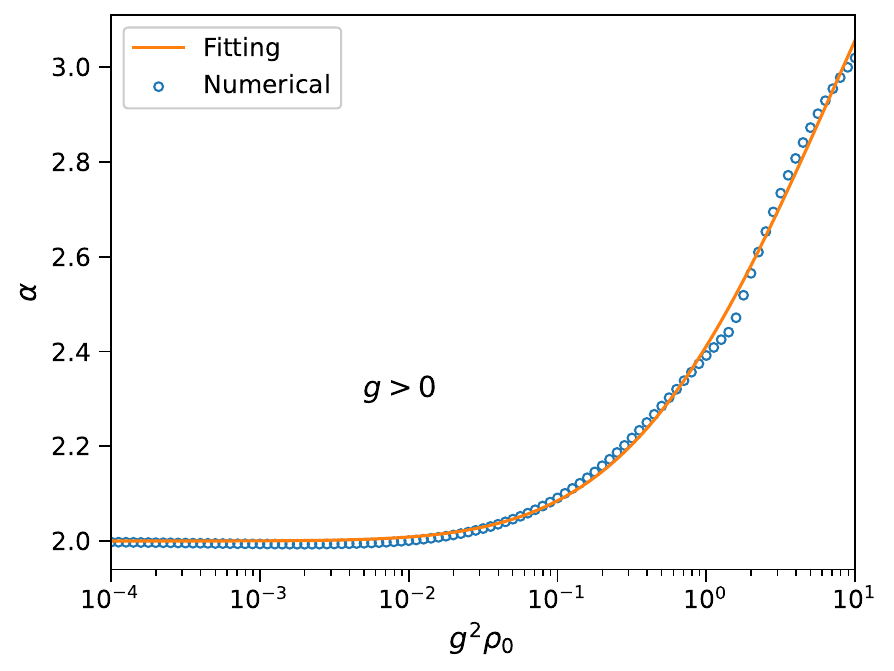}
\caption{Parameter $\alpha$ in the fitting formula Eq. (\ref{eq:sol_rho_gen}) for the case with repulsive self-interactions ($g>0$) as a function of the central density. Circles: numeric results. Solid line: fitting function, Eq. (\ref{eq:alpha}).}
\label{fig:alpha_g_pos}
\end{figure}

As in the $g<0$ case, we also fit $r^2\rho(r)$ within the range $0.01 r_{\rm core} < r < 5 r_{\rm core}$ to the
the results obtained from numerical wave function. But we have fixed $\beta$ at $8$. Allowing $\beta$ to be a free
parameter does not improve the fit too much. For the dependence of $\alpha$ on $g^2\rho$ we take the same form
as in Eq. (\ref{eq:alpha}). The best-fit $\alpha_i$ are listed in Table \ref{tab:rho_fit}. A comparison between
the fitting function of $\alpha$ and the one obtained from numerical results is shown in Fig.~\ref{fig:alpha_g_pos}
\footnote{When we derive $\alpha$ for each soliton solution by fitting Eq. (\ref{eq:sol_rho_gen}) to the numerical results, we fix the core radius using Eq. (\ref{eq:r_core_g_pos}) which is not smooth at $g^2\rho_0=1.5$. So the values of $\alpha$ with respect to $g^2\rho_0$ we obtain (circles) has a small fluctuation around that density.}.
We have fitted the soliton density for $10^{-4}<g^2\rho<10$, but we note that Eq. (\ref{eq:sol_rho_gen}) can not well describe the soliton
density at very small radii when $g^2\rho\gtrsim 1$. In those cases, the Thomas-Fermi-like solution Eq.~(\ref{eq:rho_thomas_fermi})
is more accurate at $r<r_{\rm core}$.

\begin{table}[htbp]
\begin{tabular}{ccc}
\hline
           &~~~~~~~~~~$g<0$~~~~~~~~~~&~~~~~~~~~~$g>0$~~~~~~~~~~\\
\hline
$\alpha_a$ & $1.81823$  & $3.96849$  \\
$\alpha_b$ & $1.73925$  & $2.11238$  \\
$\alpha_c$ & $0.22478$  & $0.143883$ \\
\hline
$\beta_a$  & $3.48601$  &            \\
$\beta_b$  & $1.29355$  & $\beta=8$  \\
$\beta_c$  & $0.122718$ &            \\
\hline
\end{tabular}
\caption{Best-fit parameters for the soliton density with attractive ($g<0$) and repulsive ($g>0$) self-interactions.}
\label{tab:rho_fit}
\end{table}

For other approximate soliton solutions with and without self-interactions, see \cite{Kling:2017mif,Kling:2017hjm}.

\section{Condensation time for gravitational and self-interactions}
\label{APPE:Conden_Self_Grav}
The transport cross section for self-interaction and gravity are $\sigma_{\rm self} = m^2 g^2/(2 \pi)$ and $\sigma_{\rm gravity} \sim 8{\pi}(mG)^2\Lambda/v^4$, respectively \cite{Levkov:2018kau}. The ratio of condensation time of self-interaction to  gravity can be written as
\begin{equation}
\frac{\tau_{\rm gravity}}{\tau_{\rm self}} \sim \frac{\sigma_{\rm self}}{\sigma_{\rm gravity}} 
\sim \frac{g^2v^4}{16 \pi^2 G^2 \log(mvL)},
\label{eq:time_compare}
\end{equation}
where $\tau_{\rm self}$ is the condensation time due to self-interaction. 
Using Eq.~(\ref{eq:time_compare}), we can estimate the effect of self-interaction and gravity on the condensation of boson stars.
For example, for a system of typical QCD axions with $v \approx 10^{-9}$, and decay constant $f_a \approx 10^{11} {\rm GeV}$ ($g=-\frac{1}{12 f_a^2}$), we have $\tau_{\rm gravity}/\tau_{\rm self} \ll 1$, thus gravity plays a much more important role in the condensation process.

Recently, Kay Kirkpatrick et al.~\cite{Kirkpatrick:2020fwd} argue that the relaxation rate due to self-interaction is proportional to $|g|$ rather than $g^2$, suggesting a much shorter condensation time for self-interaction compared to the one reported by other literature. Further simulations are needed to verify this. However, for typical QCD axions gravity still dominates the condensation process.

\bibliography{Reference}

\begin{thebibliography}{81}%
\makeatletter
\providecommand \@ifxundefined [1]{%
 \@ifx{#1\undefined}
}%
\providecommand \@ifnum [1]{%
 \ifnum #1\expandafter \@firstoftwo
 \else \expandafter \@secondoftwo
 \fi
}%
\providecommand \@ifx [1]{%
 \ifx #1\expandafter \@firstoftwo
 \else \expandafter \@secondoftwo
 \fi
}%
\providecommand \natexlab [1]{#1}%
\providecommand \enquote  [1]{``#1''}%
\providecommand \bibnamefont  [1]{#1}%
\providecommand \bibfnamefont [1]{#1}%
\providecommand \citenamefont [1]{#1}%
\providecommand \href@noop [0]{\@secondoftwo}%
\providecommand \href [0]{\begingroup \@sanitize@url \@href}%
\providecommand \@href[1]{\@@startlink{#1}\@@href}%
\providecommand \@@href[1]{\endgroup#1\@@endlink}%
\providecommand \@sanitize@url [0]{\catcode `\\12\catcode `\$12\catcode
  `\&12\catcode `\#12\catcode `\^12\catcode `\_12\catcode `\%12\relax}%
\providecommand \@@startlink[1]{}%
\providecommand \@@endlink[0]{}%
\providecommand \url  [0]{\begingroup\@sanitize@url \@url }%
\providecommand \@url [1]{\endgroup\@href {#1}{\urlprefix }}%
\providecommand \urlprefix  [0]{URL }%
\providecommand \Eprint [0]{\href }%
\providecommand \doibase [0]{https://doi.org/}%
\providecommand \selectlanguage [0]{\@gobble}%
\providecommand \bibinfo  [0]{\@secondoftwo}%
\providecommand \bibfield  [0]{\@secondoftwo}%
\providecommand \translation [1]{[#1]}%
\providecommand \BibitemOpen [0]{}%
\providecommand \bibitemStop [0]{}%
\providecommand \bibitemNoStop [0]{.\EOS\space}%
\providecommand \EOS [0]{\spacefactor3000\relax}%
\providecommand \BibitemShut  [1]{\csname bibitem#1\endcsname}%
\let\auto@bib@innerbib\@empty
\bibitem [{\citenamefont {Aghanim}\ \emph {et~al.}(2020)\citenamefont {Aghanim}
  \emph {et~al.}}]{Aghanim:2018eyx}%
  \BibitemOpen
  \bibfield  {author} {\bibinfo {author} {\bibfnamefont {N.}~\bibnamefont
  {Aghanim}} \emph {et~al.} (\bibinfo {collaboration} {Planck}),\ }\bibfield
  {title} {\bibinfo {title} {{Planck 2018 results. VI. Cosmological
  parameters}},\ }\href {https://doi.org/10.1051/0004-6361/201833910}
  {\bibfield  {journal} {\bibinfo  {journal} {Astron. Astrophys.}\ }\textbf
  {\bibinfo {volume} {641}},\ \bibinfo {pages} {A6} (\bibinfo {year} {2020})},\
  \Eprint {https://arxiv.org/abs/1807.06209} {arXiv:1807.06209 [astro-ph.CO]}
  \BibitemShut {NoStop}%
\bibitem [{\citenamefont {Dine}\ and\ \citenamefont
  {Fischler}(1983)}]{Dine:1982ah}%
  \BibitemOpen
  \bibfield  {author} {\bibinfo {author} {\bibfnamefont {M.}~\bibnamefont
  {Dine}}\ and\ \bibinfo {author} {\bibfnamefont {W.}~\bibnamefont
  {Fischler}},\ }\bibfield  {title} {\bibinfo {title} {{The Not So Harmless
  Axion}},\ }\href {https://doi.org/10.1016/0370-2693(83)90639-1} {\bibfield
  {journal} {\bibinfo  {journal} {Phys. Lett. B}\ }\textbf {\bibinfo {volume}
  {120}},\ \bibinfo {pages} {137} (\bibinfo {year} {1983})}\BibitemShut
  {NoStop}%
\bibitem [{\citenamefont {Suárez}\ \emph {et~al.}(2014)\citenamefont
  {Suárez}, \citenamefont {Robles},\ and\ \citenamefont
  {Matos}}]{Suarez:2013iw}%
  \BibitemOpen
  \bibfield  {author} {\bibinfo {author} {\bibfnamefont {A.}~\bibnamefont
  {Suárez}}, \bibinfo {author} {\bibfnamefont {V.~H.}\ \bibnamefont
  {Robles}},\ and\ \bibinfo {author} {\bibfnamefont {T.}~\bibnamefont
  {Matos}},\ }\bibfield  {title} {\bibinfo {title} {{A Review on the Scalar
  Field/Bose-Einstein Condensate Dark Matter Model}},\ }\href
  {https://doi.org/10.1007/978-3-319-02063-1\_9} {\bibfield  {journal}
  {\bibinfo  {journal} {Astrophys. Space Sci. Proc.}\ }\textbf {\bibinfo
  {volume} {38}},\ \bibinfo {pages} {107} (\bibinfo {year} {2014})},\ \Eprint
  {https://arxiv.org/abs/1302.0903} {arXiv:1302.0903 [astro-ph.CO]}
  \BibitemShut {NoStop}%
\bibitem [{\citenamefont {Preskill}\ \emph {et~al.}(1983)\citenamefont
  {Preskill}, \citenamefont {Wise},\ and\ \citenamefont
  {Wilczek}}]{Preskill:1982cy}%
  \BibitemOpen
  \bibfield  {author} {\bibinfo {author} {\bibfnamefont {J.}~\bibnamefont
  {Preskill}}, \bibinfo {author} {\bibfnamefont {M.~B.}\ \bibnamefont {Wise}},\
  and\ \bibinfo {author} {\bibfnamefont {F.}~\bibnamefont {Wilczek}},\
  }\bibfield  {title} {\bibinfo {title} {{Cosmology of the Invisible Axion}},\
  }\href {https://doi.org/10.1016/0370-2693(83)90637-8} {\bibfield  {journal}
  {\bibinfo  {journal} {Phys. Lett. B}\ }\textbf {\bibinfo {volume} {120}},\
  \bibinfo {pages} {127} (\bibinfo {year} {1983})}\BibitemShut {NoStop}%
\bibitem [{\citenamefont {Abbott}\ and\ \citenamefont
  {Sikivie}(1983)}]{Abbott:1982af}%
  \BibitemOpen
  \bibfield  {author} {\bibinfo {author} {\bibfnamefont {L.}~\bibnamefont
  {Abbott}}\ and\ \bibinfo {author} {\bibfnamefont {P.}~\bibnamefont
  {Sikivie}},\ }\bibfield  {title} {\bibinfo {title} {{A Cosmological Bound on
  the Invisible Axion}},\ }\href {https://doi.org/10.1016/0370-2693(83)90638-X}
  {\bibfield  {journal} {\bibinfo  {journal} {Phys. Lett. B}\ }\textbf
  {\bibinfo {volume} {120}},\ \bibinfo {pages} {133} (\bibinfo {year}
  {1983})}\BibitemShut {NoStop}%
\bibitem [{\citenamefont {Guth}\ \emph {et~al.}(2015)\citenamefont {Guth},
  \citenamefont {Hertzberg},\ and\ \citenamefont
  {Prescod-Weinstein}}]{Guth_2015}%
  \BibitemOpen
  \bibfield  {author} {\bibinfo {author} {\bibfnamefont {A.~H.}\ \bibnamefont
  {Guth}}, \bibinfo {author} {\bibfnamefont {M.~P.}\ \bibnamefont
  {Hertzberg}},\ and\ \bibinfo {author} {\bibfnamefont {C.}~\bibnamefont
  {Prescod-Weinstein}},\ }\bibfield  {title} {\bibinfo {title} {Do dark matter
  axions form a condensate with long-range correlation?},\ }\bibfield
  {journal} {\bibinfo  {journal} {Physical Review D}\ }\textbf {\bibinfo
  {volume} {92}},\ \href {https://doi.org/10.1103/physrevd.92.103513}
  {10.1103/physrevd.92.103513} (\bibinfo {year} {2015})\BibitemShut {NoStop}%
\bibitem [{\citenamefont {{Widrow}}\ and\ \citenamefont
  {{Kaiser}}(1993)}]{Widrow:1993qq}%
  \BibitemOpen
  \bibfield  {author} {\bibinfo {author} {\bibfnamefont {L.~M.}\ \bibnamefont
  {{Widrow}}}\ and\ \bibinfo {author} {\bibfnamefont {N.}~\bibnamefont
  {{Kaiser}}},\ }\bibfield  {title} {\bibinfo {title} {{Using the Schroedinger
  Equation to Simulate Collisionless Matter}},\ }\href
  {https://doi.org/10.1086/187073} {\bibfield  {journal} {\bibinfo  {journal}
  {\apjl}\ }\textbf {\bibinfo {volume} {416}},\ \bibinfo {pages} {L71}
  (\bibinfo {year} {1993})}\BibitemShut {NoStop}%
\bibitem [{\citenamefont {Uhlemann}\ \emph {et~al.}(2014)\citenamefont
  {Uhlemann}, \citenamefont {Kopp},\ and\ \citenamefont
  {Haugg}}]{Uhlemann:2014npa}%
  \BibitemOpen
  \bibfield  {author} {\bibinfo {author} {\bibfnamefont {C.}~\bibnamefont
  {Uhlemann}}, \bibinfo {author} {\bibfnamefont {M.}~\bibnamefont {Kopp}},\
  and\ \bibinfo {author} {\bibfnamefont {T.}~\bibnamefont {Haugg}},\ }\bibfield
   {title} {\bibinfo {title} {{Schr{\"o}dinger method as $N$-body double and UV
  completion of dust}},\ }\href {https://doi.org/10.1103/PhysRevD.90.023517}
  {\bibfield  {journal} {\bibinfo  {journal} {\prd}\ }\textbf {\bibinfo
  {volume} {90}},\ \bibinfo {pages} {023517} (\bibinfo {year} {2014})},\
  \Eprint {https://arxiv.org/abs/1403.5567} {arXiv:1403.5567 [astro-ph.CO]}
  \BibitemShut {NoStop}%
\bibitem [{\citenamefont {Sikivie}(2008)}]{Sikivie:2006ni}%
  \BibitemOpen
  \bibfield  {author} {\bibinfo {author} {\bibfnamefont {P.}~\bibnamefont
  {Sikivie}},\ }\bibfield  {title} {\bibinfo {title} {{Axion Cosmology}},\
  }\href {https://doi.org/10.1007/978-3-540-73518-2_2} {\bibfield  {journal}
  {\bibinfo  {journal} {Lect. Notes Phys.}\ }\textbf {\bibinfo {volume}
  {741}},\ \bibinfo {pages} {19} (\bibinfo {year} {2008})},\ \Eprint
  {https://arxiv.org/abs/astro-ph/0610440} {arXiv:astro-ph/0610440}
  \BibitemShut {NoStop}%
\bibitem [{\citenamefont {Arvanitaki}\ \emph {et~al.}(2010)\citenamefont
  {Arvanitaki}, \citenamefont {Dimopoulos}, \citenamefont {Dubovsky},
  \citenamefont {Kaloper},\ and\ \citenamefont
  {March-Russell}}]{Arvanitaki:2009fg}%
  \BibitemOpen
  \bibfield  {author} {\bibinfo {author} {\bibfnamefont {A.}~\bibnamefont
  {Arvanitaki}}, \bibinfo {author} {\bibfnamefont {S.}~\bibnamefont
  {Dimopoulos}}, \bibinfo {author} {\bibfnamefont {S.}~\bibnamefont
  {Dubovsky}}, \bibinfo {author} {\bibfnamefont {N.}~\bibnamefont {Kaloper}},\
  and\ \bibinfo {author} {\bibfnamefont {J.}~\bibnamefont {March-Russell}},\
  }\bibfield  {title} {\bibinfo {title} {{String Axiverse}},\ }\href
  {https://doi.org/10.1103/PhysRevD.81.123530} {\bibfield  {journal} {\bibinfo
  {journal} {Phys. Rev. D}\ }\textbf {\bibinfo {volume} {81}},\ \bibinfo
  {pages} {123530} (\bibinfo {year} {2010})},\ \Eprint
  {https://arxiv.org/abs/0905.4720} {arXiv:0905.4720 [hep-th]} \BibitemShut
  {NoStop}%
\bibitem [{\citenamefont {{Levkov}}\ \emph {et~al.}(2017)\citenamefont
  {{Levkov}}, \citenamefont {{Panin}},\ and\ \citenamefont
  {{Tkachev}}}]{2017PhRvL.118a1301L}%
  \BibitemOpen
  \bibfield  {author} {\bibinfo {author} {\bibfnamefont {D.~G.}\ \bibnamefont
  {{Levkov}}}, \bibinfo {author} {\bibfnamefont {A.~G.}\ \bibnamefont
  {{Panin}}},\ and\ \bibinfo {author} {\bibfnamefont {I.~I.}\ \bibnamefont
  {{Tkachev}}},\ }\bibfield  {title} {\bibinfo {title} {{Relativistic Axions
  from Collapsing Bose Stars}},\ }\href
  {https://doi.org/10.1103/PhysRevLett.118.011301} {\bibfield  {journal}
  {\bibinfo  {journal} {Physical Review Letters}\ }\textbf {\bibinfo {volume}
  {118}},\ \bibinfo {eid} {011301} (\bibinfo {year} {2017})},\ \Eprint
  {https://arxiv.org/abs/1609.03611} {arXiv:1609.03611} \BibitemShut {NoStop}%
\bibitem [{\citenamefont {Peccei}\ and\ \citenamefont
  {Quinn}(1977)}]{pecceiquinn1977}%
  \BibitemOpen
  \bibfield  {author} {\bibinfo {author} {\bibfnamefont {R.~D.}\ \bibnamefont
  {Peccei}}\ and\ \bibinfo {author} {\bibfnamefont {H.~R.}\ \bibnamefont
  {Quinn}},\ }\bibfield  {title} {\bibinfo {title} {{CP Conservation in the
  Presence of Instantons}},\ }\href
  {https://doi.org/10.1103/PhysRevLett.38.1440} {\bibfield  {journal} {\bibinfo
   {journal} {\prl}\ }\textbf {\bibinfo {volume} {38}},\ \bibinfo {pages}
  {1440} (\bibinfo {year} {1977})}\BibitemShut {NoStop}%
\bibitem [{\citenamefont {Weinberg}(1978)}]{weinberg1978}%
  \BibitemOpen
  \bibfield  {author} {\bibinfo {author} {\bibfnamefont {S.}~\bibnamefont
  {Weinberg}},\ }\bibfield  {title} {\bibinfo {title} {{A New Light Boson?}},\
  }\href {https://doi.org/10.1103/PhysRevLett.40.223} {\bibfield  {journal}
  {\bibinfo  {journal} {\prl}\ }\textbf {\bibinfo {volume} {40}},\ \bibinfo
  {pages} {223} (\bibinfo {year} {1978})}\BibitemShut {NoStop}%
\bibitem [{\citenamefont {Wilczek}(1978)}]{wilczek1978}%
  \BibitemOpen
  \bibfield  {author} {\bibinfo {author} {\bibfnamefont {F.}~\bibnamefont
  {Wilczek}},\ }\bibfield  {title} {\bibinfo {title} {{Problem of Strong p and
  t Invariance in the Presence of Instantons}},\ }\href
  {https://doi.org/10.1103/PhysRevLett.40.279} {\bibfield  {journal} {\bibinfo
  {journal} {\prl}\ }\textbf {\bibinfo {volume} {40}},\ \bibinfo {pages} {279}
  (\bibinfo {year} {1978})}\BibitemShut {NoStop}%
\bibitem [{\citenamefont {Kim}(1979)}]{Kim:1979if}%
  \BibitemOpen
  \bibfield  {author} {\bibinfo {author} {\bibfnamefont {J.~E.}\ \bibnamefont
  {Kim}},\ }\bibfield  {title} {\bibinfo {title} {{Weak Interaction Singlet and
  Strong CP Invariance}},\ }\href {https://doi.org/10.1103/PhysRevLett.43.103}
  {\bibfield  {journal} {\bibinfo  {journal} {Phys. Rev. Lett.}\ }\textbf
  {\bibinfo {volume} {43}},\ \bibinfo {pages} {103} (\bibinfo {year}
  {1979})}\BibitemShut {NoStop}%
\bibitem [{\citenamefont {Shifman}\ \emph {et~al.}(1980)\citenamefont
  {Shifman}, \citenamefont {Vainshtein},\ and\ \citenamefont
  {Zakharov}}]{Shifman:1979if}%
  \BibitemOpen
  \bibfield  {author} {\bibinfo {author} {\bibfnamefont {M.~A.}\ \bibnamefont
  {Shifman}}, \bibinfo {author} {\bibfnamefont {A.}~\bibnamefont
  {Vainshtein}},\ and\ \bibinfo {author} {\bibfnamefont {V.~I.}\ \bibnamefont
  {Zakharov}},\ }\bibfield  {title} {\bibinfo {title} {{Can Confinement Ensure
  Natural CP Invariance of Strong Interactions?}},\ }\href
  {https://doi.org/10.1016/0550-3213(80)90209-6} {\bibfield  {journal}
  {\bibinfo  {journal} {Nucl. Phys. B}\ }\textbf {\bibinfo {volume} {166}},\
  \bibinfo {pages} {493} (\bibinfo {year} {1980})}\BibitemShut {NoStop}%
\bibitem [{\citenamefont {Zhitnitsky}(1980)}]{Zhitnitsky:1980tq}%
  \BibitemOpen
  \bibfield  {author} {\bibinfo {author} {\bibfnamefont {A.}~\bibnamefont
  {Zhitnitsky}},\ }\bibfield  {title} {\bibinfo {title} {{On Possible
  Suppression of the Axion Hadron Interactions. (In Russian)}},\ }\href@noop {}
  {\bibfield  {journal} {\bibinfo  {journal} {Sov.J . Nucl. Phys.}\ }\textbf
  {\bibinfo {volume} {31}},\ \bibinfo {pages} {260} (\bibinfo {year}
  {1980})}\BibitemShut {NoStop}%
\bibitem [{\citenamefont {{Dine}}\ \emph {et~al.}(1981)\citenamefont {{Dine}},
  \citenamefont {{Fischler}},\ and\ \citenamefont
  {{Srednicki}}}]{1981PhLB..104..199D}%
  \BibitemOpen
  \bibfield  {author} {\bibinfo {author} {\bibfnamefont {M.}~\bibnamefont
  {{Dine}}}, \bibinfo {author} {\bibfnamefont {W.}~\bibnamefont {{Fischler}}},\
  and\ \bibinfo {author} {\bibfnamefont {M.}~\bibnamefont {{Srednicki}}},\
  }\bibfield  {title} {\bibinfo {title} {{A simple solution to the strong CP
  problem with a harmless axion}},\ }\href
  {https://doi.org/10.1016/0370-2693(81)90590-6} {\bibfield  {journal}
  {\bibinfo  {journal} {Phys. Lett. B}\ }\textbf {\bibinfo {volume} {104}},\
  \bibinfo {pages} {199} (\bibinfo {year} {1981})}\BibitemShut {NoStop}%
\bibitem [{\citenamefont {{Arvanitaki}}\ \emph {et~al.}(2015)\citenamefont
  {{Arvanitaki}}, \citenamefont {{Baryakhtar}},\ and\ \citenamefont
  {{Huang}}}]{2015PhRvD..91h4011A}%
  \BibitemOpen
  \bibfield  {author} {\bibinfo {author} {\bibfnamefont {A.}~\bibnamefont
  {{Arvanitaki}}}, \bibinfo {author} {\bibfnamefont {M.}~\bibnamefont
  {{Baryakhtar}}},\ and\ \bibinfo {author} {\bibfnamefont {X.}~\bibnamefont
  {{Huang}}},\ }\bibfield  {title} {\bibinfo {title} {{Discovering the QCD
  axion with black holes and gravitational waves}},\ }\href
  {https://doi.org/10.1103/PhysRevD.91.084011} {\bibfield  {journal} {\bibinfo
  {journal} {\prd}\ }\textbf {\bibinfo {volume} {91}},\ \bibinfo {eid} {084011}
  (\bibinfo {year} {2015})},\ \Eprint {https://arxiv.org/abs/1411.2263}
  {arXiv:1411.2263 [hep-ph]} \BibitemShut {NoStop}%
\bibitem [{\citenamefont {{Payez}}\ \emph {et~al.}(2015)\citenamefont
  {{Payez}}, \citenamefont {{Evoli}}, \citenamefont {{Fischer}}, \citenamefont
  {{Giannotti}}, \citenamefont {{Mirizzi}},\ and\ \citenamefont
  {{Ringwald}}}]{2015JCAP...02..006P}%
  \BibitemOpen
  \bibfield  {author} {\bibinfo {author} {\bibfnamefont {A.}~\bibnamefont
  {{Payez}}}, \bibinfo {author} {\bibfnamefont {C.}~\bibnamefont {{Evoli}}},
  \bibinfo {author} {\bibfnamefont {T.}~\bibnamefont {{Fischer}}}, \bibinfo
  {author} {\bibfnamefont {M.}~\bibnamefont {{Giannotti}}}, \bibinfo {author}
  {\bibfnamefont {A.}~\bibnamefont {{Mirizzi}}},\ and\ \bibinfo {author}
  {\bibfnamefont {A.}~\bibnamefont {{Ringwald}}},\ }\bibfield  {title}
  {\bibinfo {title} {{Revisiting the SN1987A gamma-ray limit on ultralight
  axion-like particles}},\ }\href
  {https://doi.org/10.1088/1475-7516/2015/02/006} {\bibfield  {journal}
  {\bibinfo  {journal} {\jcap}\ }\textbf {\bibinfo {volume} {2}},\ \bibinfo
  {eid} {006} (\bibinfo {year} {2015})},\ \Eprint
  {https://arxiv.org/abs/1410.3747} {arXiv:1410.3747 [astro-ph.HE]}
  \BibitemShut {NoStop}%
\bibitem [{\citenamefont {{Marsh}}(2016)}]{Marsh:2015xka}%
  \BibitemOpen
  \bibfield  {author} {\bibinfo {author} {\bibfnamefont {D.~J.~E.}\
  \bibnamefont {{Marsh}}},\ }\bibfield  {title} {\bibinfo {title} {{Axion
  cosmology}},\ }\href {https://doi.org/10.1016/j.physrep.2016.06.005}
  {\bibfield  {journal} {\bibinfo  {journal} {\physrep}\ }\textbf {\bibinfo
  {volume} {643}},\ \bibinfo {pages} {1} (\bibinfo {year} {2016})},\ \Eprint
  {https://arxiv.org/abs/1510.07633} {arXiv:1510.07633} \BibitemShut {NoStop}%
\bibitem [{\citenamefont {Tanabashi}\ \emph {et~al.}(2018)\citenamefont
  {Tanabashi} \emph {et~al.}}]{Tanabashi:2018oca}%
  \BibitemOpen
  \bibfield  {author} {\bibinfo {author} {\bibfnamefont {M.}~\bibnamefont
  {Tanabashi}} \emph {et~al.} (\bibinfo {collaboration} {Particle Data
  Group}),\ }\bibfield  {title} {\bibinfo {title} {{Review of Particle
  Physics}},\ }\href {https://doi.org/10.1103/PhysRevD.98.030001} {\bibfield
  {journal} {\bibinfo  {journal} {Phys. Rev. D}\ }\textbf {\bibinfo {volume}
  {98}},\ \bibinfo {pages} {030001} (\bibinfo {year} {2018})}\BibitemShut
  {NoStop}%
\bibitem [{\citenamefont {Luzio}\ \emph {et~al.}(2020)\citenamefont {Luzio},
  \citenamefont {Giannotti}, \citenamefont {Nardi},\ and\ \citenamefont
  {Visinelli}}]{luzio2020landscape}%
  \BibitemOpen
  \bibfield  {author} {\bibinfo {author} {\bibfnamefont {L.~D.}\ \bibnamefont
  {Luzio}}, \bibinfo {author} {\bibfnamefont {M.}~\bibnamefont {Giannotti}},
  \bibinfo {author} {\bibfnamefont {E.}~\bibnamefont {Nardi}},\ and\ \bibinfo
  {author} {\bibfnamefont {L.}~\bibnamefont {Visinelli}},\ }\href@noop {}
  {\bibinfo {title} {The landscape of qcd axion models}} (\bibinfo {year}
  {2020}),\ \Eprint {https://arxiv.org/abs/2003.01100} {arXiv:2003.01100
  [hep-ph]} \BibitemShut {NoStop}%
\bibitem [{\citenamefont {{Press}}\ \emph {et~al.}(1990)\citenamefont
  {{Press}}, \citenamefont {{Ryden}},\ and\ \citenamefont
  {{Spergel}}}]{1990PhRvL..64.1084P}%
  \BibitemOpen
  \bibfield  {author} {\bibinfo {author} {\bibfnamefont {W.~H.}\ \bibnamefont
  {{Press}}}, \bibinfo {author} {\bibfnamefont {B.~S.}\ \bibnamefont
  {{Ryden}}},\ and\ \bibinfo {author} {\bibfnamefont {D.~N.}\ \bibnamefont
  {{Spergel}}},\ }\bibfield  {title} {\bibinfo {title} {{Single mechanism for
  generating large-scale structure and providing dark missing matter}},\ }\href
  {https://doi.org/10.1103/PhysRevLett.64.1084} {\bibfield  {journal} {\bibinfo
   {journal} {\prl}\ }\textbf {\bibinfo {volume} {64}},\ \bibinfo {pages}
  {1084} (\bibinfo {year} {1990})}\BibitemShut {NoStop}%
\bibitem [{\citenamefont {{Sahni}}\ and\ \citenamefont
  {{Wang}}(2000)}]{2000PhRvD..62j3517S}%
  \BibitemOpen
  \bibfield  {author} {\bibinfo {author} {\bibfnamefont {V.}~\bibnamefont
  {{Sahni}}}\ and\ \bibinfo {author} {\bibfnamefont {L.}~\bibnamefont
  {{Wang}}},\ }\bibfield  {title} {\bibinfo {title} {{New cosmological model of
  quintessence and dark matter}},\ }\href
  {https://doi.org/10.1103/PhysRevD.62.103517} {\bibfield  {journal} {\bibinfo
  {journal} {\prd}\ }\textbf {\bibinfo {volume} {62}},\ \bibinfo {eid} {103517}
  (\bibinfo {year} {2000})},\ \Eprint {https://arxiv.org/abs/astro-ph/9910097}
  {astro-ph/9910097} \BibitemShut {NoStop}%
\bibitem [{\citenamefont {Hu}\ \emph {et~al.}(2000)\citenamefont {Hu},
  \citenamefont {Barkana},\ and\ \citenamefont {Gruzinov}}]{hu2000}%
  \BibitemOpen
  \bibfield  {author} {\bibinfo {author} {\bibfnamefont {W.}~\bibnamefont
  {Hu}}, \bibinfo {author} {\bibfnamefont {R.}~\bibnamefont {Barkana}},\ and\
  \bibinfo {author} {\bibfnamefont {A.}~\bibnamefont {Gruzinov}},\ }\bibfield
  {title} {\bibinfo {title} {{Cold and fuzzy dark matter}},\ }\href
  {https://doi.org/10.1103/PhysRevLett.85.1158} {\bibfield  {journal} {\bibinfo
   {journal} {\prl}\ }\textbf {\bibinfo {volume} {85}},\ \bibinfo {pages}
  {1158} (\bibinfo {year} {2000})},\ \Eprint
  {https://arxiv.org/abs/astro-ph/0003365} {astro-ph/0003365} \BibitemShut
  {NoStop}%
\bibitem [{\citenamefont {{Peebles}}(2000)}]{2000ApJ...534L.127P}%
  \BibitemOpen
  \bibfield  {author} {\bibinfo {author} {\bibfnamefont {P.~J.~E.}\
  \bibnamefont {{Peebles}}},\ }\bibfield  {title} {\bibinfo {title} {{Fluid
  Dark Matter}},\ }\href {https://doi.org/10.1086/312677} {\bibfield  {journal}
  {\bibinfo  {journal} {\apjl}\ }\textbf {\bibinfo {volume} {534}},\ \bibinfo
  {pages} {L127} (\bibinfo {year} {2000})},\ \Eprint
  {https://arxiv.org/abs/astro-ph/0002495} {astro-ph/0002495} \BibitemShut
  {NoStop}%
\bibitem [{\citenamefont {{Marsh}}\ and\ \citenamefont
  {{Pop}}(2015)}]{2015MNRAS.451.2479M}%
  \BibitemOpen
  \bibfield  {author} {\bibinfo {author} {\bibfnamefont {D.~J.~E.}\
  \bibnamefont {{Marsh}}}\ and\ \bibinfo {author} {\bibfnamefont {A.-R.}\
  \bibnamefont {{Pop}}},\ }\bibfield  {title} {\bibinfo {title} {{Axion dark
  matter, solitons and the cusp-core problem}},\ }\href
  {https://doi.org/10.1093/mnras/stv1050} {\bibfield  {journal} {\bibinfo
  {journal} {\mnras}\ }\textbf {\bibinfo {volume} {451}},\ \bibinfo {pages}
  {2479} (\bibinfo {year} {2015})},\ \Eprint {https://arxiv.org/abs/1502.03456}
  {arXiv:1502.03456} \BibitemShut {NoStop}%
\bibitem [{\citenamefont {Hui}\ \emph {et~al.}(2017)\citenamefont {Hui},
  \citenamefont {Ostriker}, \citenamefont {Tremaine},\ and\ \citenamefont
  {Witten}}]{Hui:2016ltb}%
  \BibitemOpen
  \bibfield  {author} {\bibinfo {author} {\bibfnamefont {L.}~\bibnamefont
  {Hui}}, \bibinfo {author} {\bibfnamefont {J.~P.}\ \bibnamefont {Ostriker}},
  \bibinfo {author} {\bibfnamefont {S.}~\bibnamefont {Tremaine}},\ and\
  \bibinfo {author} {\bibfnamefont {E.}~\bibnamefont {Witten}},\ }\bibfield
  {title} {\bibinfo {title} {{Ultralight scalars as cosmological dark
  matter}},\ }\href {https://doi.org/10.1103/PhysRevD.95.043541} {\bibfield
  {journal} {\bibinfo  {journal} {Phys. Rev.}\ }\textbf {\bibinfo {volume}
  {D95}},\ \bibinfo {pages} {043541} (\bibinfo {year} {2017})},\ \Eprint
  {https://arxiv.org/abs/1610.08297} {arXiv:1610.08297 [astro-ph.CO]}
  \BibitemShut {NoStop}%
\bibitem [{\citenamefont {Seidel}\ and\ \citenamefont
  {Suen}(1990)}]{PhysRevD.42.384}%
  \BibitemOpen
  \bibfield  {author} {\bibinfo {author} {\bibfnamefont {E.}~\bibnamefont
  {Seidel}}\ and\ \bibinfo {author} {\bibfnamefont {W.-M.}\ \bibnamefont
  {Suen}},\ }\bibfield  {title} {\bibinfo {title} {Dynamical evolution of boson
  stars: Perturbing the ground state},\ }\href
  {https://doi.org/10.1103/PhysRevD.42.384} {\bibfield  {journal} {\bibinfo
  {journal} {Phys. Rev. D}\ }\textbf {\bibinfo {volume} {42}},\ \bibinfo
  {pages} {384} (\bibinfo {year} {1990})}\BibitemShut {NoStop}%
\bibitem [{\citenamefont {{Kaup}}(1968)}]{1968PhRv..172.1331K}%
  \BibitemOpen
  \bibfield  {author} {\bibinfo {author} {\bibfnamefont {D.~J.}\ \bibnamefont
  {{Kaup}}},\ }\bibfield  {title} {\bibinfo {title} {{Klein-Gordon Geon}},\
  }\href {https://doi.org/10.1103/PhysRev.172.1331} {\bibfield  {journal}
  {\bibinfo  {journal} {Physical Review}\ }\textbf {\bibinfo {volume} {172}},\
  \bibinfo {pages} {1331} (\bibinfo {year} {1968})}\BibitemShut {NoStop}%
\bibitem [{\citenamefont {{Ruffini}}\ and\ \citenamefont
  {{Bonazzola}}(1969)}]{1969PhRv..187.1767R}%
  \BibitemOpen
  \bibfield  {author} {\bibinfo {author} {\bibfnamefont {R.}~\bibnamefont
  {{Ruffini}}}\ and\ \bibinfo {author} {\bibfnamefont {S.}~\bibnamefont
  {{Bonazzola}}},\ }\bibfield  {title} {\bibinfo {title} {{Systems of
  Self-Gravitating Particles in General Relativity and the Concept of an
  Equation of State}},\ }\href {https://doi.org/10.1103/PhysRev.187.1767}
  {\bibfield  {journal} {\bibinfo  {journal} {Physical Review}\ }\textbf
  {\bibinfo {volume} {187}},\ \bibinfo {pages} {1767} (\bibinfo {year}
  {1969})}\BibitemShut {NoStop}%
\bibitem [{\citenamefont {{Seidel}}\ and\ \citenamefont
  {{Suen}}(1994)}]{1994PhRvL..72.2516S}%
  \BibitemOpen
  \bibfield  {author} {\bibinfo {author} {\bibfnamefont {E.}~\bibnamefont
  {{Seidel}}}\ and\ \bibinfo {author} {\bibfnamefont {W.-M.}\ \bibnamefont
  {{Suen}}},\ }\bibfield  {title} {\bibinfo {title} {{Formation of solitonic
  stars through gravitational cooling}},\ }\href
  {https://doi.org/10.1103/PhysRevLett.72.2516} {\bibfield  {journal} {\bibinfo
   {journal} {\prl}\ }\textbf {\bibinfo {volume} {72}},\ \bibinfo {pages}
  {2516} (\bibinfo {year} {1994})},\ \Eprint
  {https://arxiv.org/abs/gr-qc/9309015} {gr-qc/9309015} \BibitemShut {NoStop}%
\bibitem [{\citenamefont {Alcubierre}\ \emph {et~al.}(2003)\citenamefont
  {Alcubierre}, \citenamefont {Becerril}, \citenamefont {Guzman}, \citenamefont
  {Matos}, \citenamefont {Nunez},\ and\ \citenamefont
  {Urena-Lopez}}]{Alcubierre:2003sx}%
  \BibitemOpen
  \bibfield  {author} {\bibinfo {author} {\bibfnamefont {M.}~\bibnamefont
  {Alcubierre}}, \bibinfo {author} {\bibfnamefont {R.}~\bibnamefont
  {Becerril}}, \bibinfo {author} {\bibfnamefont {S.~F.}\ \bibnamefont
  {Guzman}}, \bibinfo {author} {\bibfnamefont {T.}~\bibnamefont {Matos}},
  \bibinfo {author} {\bibfnamefont {D.}~\bibnamefont {Nunez}},\ and\ \bibinfo
  {author} {\bibfnamefont {L.~A.}\ \bibnamefont {Urena-Lopez}},\ }\bibfield
  {title} {\bibinfo {title} {{Numerical studies of Phi**2 oscillatons}},\
  }\href {https://doi.org/10.1088/0264-9381/20/13/332} {\bibfield  {journal}
  {\bibinfo  {journal} {Class. Quant. Grav.}\ }\textbf {\bibinfo {volume}
  {20}},\ \bibinfo {pages} {2883} (\bibinfo {year} {2003})},\ \Eprint
  {https://arxiv.org/abs/gr-qc/0301105} {arXiv:gr-qc/0301105 [gr-qc]}
  \BibitemShut {NoStop}%
\bibitem [{\citenamefont {{Helfer}}\ \emph {et~al.}(2017)\citenamefont
  {{Helfer}}, \citenamefont {{Marsh}}, \citenamefont {{Clough}}, \citenamefont
  {{Fairbairn}}, \citenamefont {{Lim}},\ and\ \citenamefont
  {{Becerril}}}]{2017JCAP...03..055H}%
  \BibitemOpen
  \bibfield  {author} {\bibinfo {author} {\bibfnamefont {T.}~\bibnamefont
  {{Helfer}}}, \bibinfo {author} {\bibfnamefont {D.~J.~E.}\ \bibnamefont
  {{Marsh}}}, \bibinfo {author} {\bibfnamefont {K.}~\bibnamefont {{Clough}}},
  \bibinfo {author} {\bibfnamefont {M.}~\bibnamefont {{Fairbairn}}}, \bibinfo
  {author} {\bibfnamefont {E.~A.}\ \bibnamefont {{Lim}}},\ and\ \bibinfo
  {author} {\bibfnamefont {R.}~\bibnamefont {{Becerril}}},\ }\bibfield  {title}
  {\bibinfo {title} {{Black hole formation from axion stars}},\ }\href
  {https://doi.org/10.1088/1475-7516/2017/03/055} {\bibfield  {journal}
  {\bibinfo  {journal} {\jcap}\ }\textbf {\bibinfo {volume} {3}},\ \bibinfo
  {eid} {055} (\bibinfo {year} {2017})},\ \Eprint
  {https://arxiv.org/abs/1609.04724} {arXiv:1609.04724} \BibitemShut {NoStop}%
\bibitem [{\citenamefont {Liddle}\ and\ \citenamefont
  {Madsen}(1992)}]{Liddle:1993ha}%
  \BibitemOpen
  \bibfield  {author} {\bibinfo {author} {\bibfnamefont {A.~R.}\ \bibnamefont
  {Liddle}}\ and\ \bibinfo {author} {\bibfnamefont {M.~S.}\ \bibnamefont
  {Madsen}},\ }\bibfield  {title} {\bibinfo {title} {{The Structure and
  formation of boson stars}},\ }\href
  {https://doi.org/10.1142/S0218271892000057} {\bibfield  {journal} {\bibinfo
  {journal} {Int.J.Mod.Phys.}\ }\textbf {\bibinfo {volume} {D1}},\ \bibinfo
  {pages} {101} (\bibinfo {year} {1992})}\BibitemShut {NoStop}%
\bibitem [{\citenamefont {{Schive}}\ \emph
  {et~al.}(2014{\natexlab{a}})\citenamefont {{Schive}}, \citenamefont
  {{Chiueh}},\ and\ \citenamefont {{Broadhurst}}}]{Schive:2014dra}%
  \BibitemOpen
  \bibfield  {author} {\bibinfo {author} {\bibfnamefont {H.-Y.}\ \bibnamefont
  {{Schive}}}, \bibinfo {author} {\bibfnamefont {T.}~\bibnamefont {{Chiueh}}},\
  and\ \bibinfo {author} {\bibfnamefont {T.}~\bibnamefont {{Broadhurst}}},\
  }\bibfield  {title} {\bibinfo {title} {{Cosmic structure as the quantum
  interference of a coherent dark wave}},\ }\href
  {https://doi.org/10.1038/nphys2996} {\bibfield  {journal} {\bibinfo
  {journal} {Nature Physics}\ }\textbf {\bibinfo {volume} {10}},\ \bibinfo
  {pages} {496} (\bibinfo {year} {2014}{\natexlab{a}})},\ \Eprint
  {https://arxiv.org/abs/1406.6586} {arXiv:1406.6586} \BibitemShut {NoStop}%
\bibitem [{\citenamefont {{Schive}}\ \emph
  {et~al.}(2014{\natexlab{b}})\citenamefont {{Schive}}, \citenamefont {{Liao}},
  \citenamefont {{Woo}}, \citenamefont {{Wong}}, \citenamefont {{Chiueh}},
  \citenamefont {{Broadhurst}},\ and\ \citenamefont
  {{Hwang}}}]{Schive:2014hza}%
  \BibitemOpen
  \bibfield  {author} {\bibinfo {author} {\bibfnamefont {H.-Y.}\ \bibnamefont
  {{Schive}}}, \bibinfo {author} {\bibfnamefont {M.-H.}\ \bibnamefont
  {{Liao}}}, \bibinfo {author} {\bibfnamefont {T.-P.}\ \bibnamefont {{Woo}}},
  \bibinfo {author} {\bibfnamefont {S.-K.}\ \bibnamefont {{Wong}}}, \bibinfo
  {author} {\bibfnamefont {T.}~\bibnamefont {{Chiueh}}}, \bibinfo {author}
  {\bibfnamefont {T.}~\bibnamefont {{Broadhurst}}},\ and\ \bibinfo {author}
  {\bibfnamefont {W.-Y.~P.}\ \bibnamefont {{Hwang}}},\ }\bibfield  {title}
  {\bibinfo {title} {{Understanding the Core-Halo Relation of Quantum Wave Dark
  Matter from 3D Simulations}},\ }\href
  {https://doi.org/10.1103/PhysRevLett.113.261302} {\bibfield  {journal}
  {\bibinfo  {journal} {\prl}\ }\textbf {\bibinfo {volume} {113}},\ \bibinfo
  {eid} {261302} (\bibinfo {year} {2014}{\natexlab{b}})},\ \Eprint
  {https://arxiv.org/abs/1407.7762} {arXiv:1407.7762} \BibitemShut {NoStop}%
\bibitem [{\citenamefont {Schwabe}\ \emph {et~al.}(2016)\citenamefont
  {Schwabe}, \citenamefont {Niemeyer},\ and\ \citenamefont
  {Engels}}]{Schwabe:2016rze}%
  \BibitemOpen
  \bibfield  {author} {\bibinfo {author} {\bibfnamefont {B.}~\bibnamefont
  {Schwabe}}, \bibinfo {author} {\bibfnamefont {J.~C.}\ \bibnamefont
  {Niemeyer}},\ and\ \bibinfo {author} {\bibfnamefont {J.~F.}\ \bibnamefont
  {Engels}},\ }\bibfield  {title} {\bibinfo {title} {{Simulations of solitonic
  core mergers in ultralight axion dark matter cosmologies}},\ }\href
  {https://doi.org/10.1103/PhysRevD.94.043513} {\bibfield  {journal} {\bibinfo
  {journal} {Phys. Rev.}\ }\textbf {\bibinfo {volume} {D94}},\ \bibinfo {pages}
  {043513} (\bibinfo {year} {2016})},\ \Eprint
  {https://arxiv.org/abs/1606.05151} {arXiv:1606.05151 [astro-ph.CO]}
  \BibitemShut {NoStop}%
\bibitem [{\citenamefont {Levkov}\ \emph {et~al.}(2018)\citenamefont {Levkov},
  \citenamefont {Panin},\ and\ \citenamefont {Tkachev}}]{Levkov:2018kau}%
  \BibitemOpen
  \bibfield  {author} {\bibinfo {author} {\bibfnamefont {D.~G.}\ \bibnamefont
  {Levkov}}, \bibinfo {author} {\bibfnamefont {A.~G.}\ \bibnamefont {Panin}},\
  and\ \bibinfo {author} {\bibfnamefont {I.~I.}\ \bibnamefont {Tkachev}},\
  }\bibfield  {title} {\bibinfo {title} {{Gravitational Bose-Einstein
  condensation in the kinetic regime}},\ }\href
  {https://doi.org/10.1103/PhysRevLett.121.151301} {\bibfield  {journal}
  {\bibinfo  {journal} {Phys. Rev. Lett.}\ }\textbf {\bibinfo {volume} {121}},\
  \bibinfo {pages} {151301} (\bibinfo {year} {2018})},\ \Eprint
  {https://arxiv.org/abs/1804.05857} {arXiv:1804.05857 [astro-ph.CO]}
  \BibitemShut {NoStop}%
\bibitem [{\citenamefont {Widdicombe}\ \emph {et~al.}(2018)\citenamefont
  {Widdicombe}, \citenamefont {Helfer}, \citenamefont {Marsh},\ and\
  \citenamefont {Lim}}]{Widdicombe:2018oeo}%
  \BibitemOpen
  \bibfield  {author} {\bibinfo {author} {\bibfnamefont {J.~Y.}\ \bibnamefont
  {Widdicombe}}, \bibinfo {author} {\bibfnamefont {T.}~\bibnamefont {Helfer}},
  \bibinfo {author} {\bibfnamefont {D.~J.~E.}\ \bibnamefont {Marsh}},\ and\
  \bibinfo {author} {\bibfnamefont {E.~A.}\ \bibnamefont {Lim}},\ }\bibfield
  {title} {\bibinfo {title} {{Formation of Relativistic Axion Stars}},\ }\href
  {https://doi.org/10.1088/1475-7516/2018/10/005} {\bibfield  {journal}
  {\bibinfo  {journal} {JCAP}\ }\textbf {\bibinfo {volume} {1810}}\bibfield
  {number} {\bibinfo  {number} { (10)},\ \bibinfo {pages} {005}},\ }\Eprint
  {https://arxiv.org/abs/1806.09367} {arXiv:1806.09367 [astro-ph.CO]}
  \BibitemShut {NoStop}%
\bibitem [{\citenamefont {Eggemeier}\ and\ \citenamefont
  {Niemeyer}(2019)}]{Eggemeier:2019jsu}%
  \BibitemOpen
  \bibfield  {author} {\bibinfo {author} {\bibfnamefont {B.}~\bibnamefont
  {Eggemeier}}\ and\ \bibinfo {author} {\bibfnamefont {J.~C.}\ \bibnamefont
  {Niemeyer}},\ }\bibfield  {title} {\bibinfo {title} {{Formation and mass
  growth of axion stars in axion miniclusters}},\ }\href
  {https://doi.org/10.1103/PhysRevD.100.063528} {\bibfield  {journal} {\bibinfo
   {journal} {Phys. Rev. D}\ }\textbf {\bibinfo {volume} {100}},\ \bibinfo
  {pages} {063528} (\bibinfo {year} {2019})},\ \Eprint
  {https://arxiv.org/abs/1906.01348} {arXiv:1906.01348 [astro-ph.CO]}
  \BibitemShut {NoStop}%
\bibitem [{\citenamefont {{Bird}}\ \emph {et~al.}(2016)\citenamefont {{Bird}},
  \citenamefont {{Cholis}}, \citenamefont {{Mu{\~n}oz}}, \citenamefont
  {{Ali-Ha{\"\i}moud}}, \citenamefont {{Kamionkowski}}, \citenamefont
  {{Kovetz}}, \citenamefont {{Raccanelli}},\ and\ \citenamefont
  {{Riess}}}]{2016PhRvL.116t1301B}%
  \BibitemOpen
  \bibfield  {author} {\bibinfo {author} {\bibfnamefont {S.}~\bibnamefont
  {{Bird}}}, \bibinfo {author} {\bibfnamefont {I.}~\bibnamefont {{Cholis}}},
  \bibinfo {author} {\bibfnamefont {J.~B.}\ \bibnamefont {{Mu{\~n}oz}}},
  \bibinfo {author} {\bibfnamefont {Y.}~\bibnamefont {{Ali-Ha{\"\i}moud}}},
  \bibinfo {author} {\bibfnamefont {M.}~\bibnamefont {{Kamionkowski}}},
  \bibinfo {author} {\bibfnamefont {E.~D.}\ \bibnamefont {{Kovetz}}}, \bibinfo
  {author} {\bibfnamefont {A.}~\bibnamefont {{Raccanelli}}},\ and\ \bibinfo
  {author} {\bibfnamefont {A.~G.}\ \bibnamefont {{Riess}}},\ }\bibfield
  {title} {\bibinfo {title} {{Did LIGO Detect Dark Matter?}},\ }\href
  {https://doi.org/10.1103/PhysRevLett.116.201301} {\bibfield  {journal}
  {\bibinfo  {journal} {\prl}\ }\textbf {\bibinfo {volume} {116}},\ \bibinfo
  {eid} {201301} (\bibinfo {year} {2016})},\ \Eprint
  {https://arxiv.org/abs/1603.00464} {arXiv:1603.00464 [astro-ph.CO]}
  \BibitemShut {NoStop}%
\bibitem [{\citenamefont {Wu}\ \emph {et~al.}(1998)\citenamefont {Wu},
  \citenamefont {Chiueh}, \citenamefont {Fang},\ and\ \citenamefont
  {Xue}}]{Wu:1998ju}%
  \BibitemOpen
  \bibfield  {author} {\bibinfo {author} {\bibfnamefont {X.-P.}\ \bibnamefont
  {Wu}}, \bibinfo {author} {\bibfnamefont {T.}~\bibnamefont {Chiueh}}, \bibinfo
  {author} {\bibfnamefont {L.-Z.}\ \bibnamefont {Fang}},\ and\ \bibinfo
  {author} {\bibfnamefont {Y.-J.}\ \bibnamefont {Xue}},\ }\bibfield  {title}
  {\bibinfo {title} {{A comparison of different cluster mass estimates:
  consistency or discrepancy ?}},\ }\href
  {https://doi.org/10.1046/j.1365-8711.1998.02055.x} {\bibfield  {journal}
  {\bibinfo  {journal} {Mon. Not. Roy. Astron. Soc.}\ }\textbf {\bibinfo
  {volume} {301}},\ \bibinfo {pages} {861} (\bibinfo {year} {1998})},\ \Eprint
  {https://arxiv.org/abs/astro-ph/9808179} {arXiv:astro-ph/9808179 [astro-ph]}
  \BibitemShut {NoStop}%
\bibitem [{\citenamefont {Hinshaw}\ \emph {et~al.}(2009)\citenamefont {Hinshaw}
  \emph {et~al.}}]{Hinshaw:2008kr}%
  \BibitemOpen
  \bibfield  {author} {\bibinfo {author} {\bibfnamefont {G.}~\bibnamefont
  {Hinshaw}} \emph {et~al.} (\bibinfo {collaboration} {WMAP}),\ }\bibfield
  {title} {\bibinfo {title} {{Five-Year Wilkinson Microwave Anisotropy Probe
  (WMAP) Observations: Data Processing, Sky Maps, and Basic Results}},\ }\href
  {https://doi.org/10.1088/0067-0049/180/2/225} {\bibfield  {journal} {\bibinfo
   {journal} {Astrophys. J. Suppl.}\ }\textbf {\bibinfo {volume} {180}},\
  \bibinfo {pages} {225} (\bibinfo {year} {2009})},\ \Eprint
  {https://arxiv.org/abs/0803.0732} {arXiv:0803.0732 [astro-ph]} \BibitemShut
  {NoStop}%
\bibitem [{\citenamefont {Natarajan}\ \emph {et~al.}(2017)\citenamefont
  {Natarajan} \emph {et~al.}}]{Natarajan:2017sbo}%
  \BibitemOpen
  \bibfield  {author} {\bibinfo {author} {\bibfnamefont {P.}~\bibnamefont
  {Natarajan}} \emph {et~al.},\ }\bibfield  {title} {\bibinfo {title} {{Mapping
  substructure in the HST Frontier Fields cluster lenses and in cosmological
  simulations}},\ }\href {https://doi.org/10.1093/mnras/stw3385} {\bibfield
  {journal} {\bibinfo  {journal} {Mon. Not. Roy. Astron. Soc.}\ }\textbf
  {\bibinfo {volume} {468}},\ \bibinfo {pages} {1962} (\bibinfo {year}
  {2017})},\ \Eprint {https://arxiv.org/abs/1702.04348} {arXiv:1702.04348
  [astro-ph.GA]} \BibitemShut {NoStop}%
\bibitem [{\citenamefont {{Refregier}}(2003)}]{2003ARA&A..41..645R}%
  \BibitemOpen
  \bibfield  {author} {\bibinfo {author} {\bibfnamefont {A.}~\bibnamefont
  {{Refregier}}},\ }\bibfield  {title} {\bibinfo {title} {{Weak Gravitational
  Lensing by Large-Scale Structure}},\ }\href
  {https://doi.org/10.1146/annurev.astro.41.111302.102207} {\bibfield
  {journal} {\bibinfo  {journal} {\araa}\ }\textbf {\bibinfo {volume} {41}},\
  \bibinfo {pages} {645} (\bibinfo {year} {2003})},\ \Eprint
  {https://arxiv.org/abs/astro-ph/0307212} {arXiv:astro-ph/0307212 [astro-ph]}
  \BibitemShut {NoStop}%
\bibitem [{\citenamefont {{Bertone}}\ and\ \citenamefont
  {{Merritt}}(2005)}]{2005MPLA...20.1021B}%
  \BibitemOpen
  \bibfield  {author} {\bibinfo {author} {\bibfnamefont {G.}~\bibnamefont
  {{Bertone}}}\ and\ \bibinfo {author} {\bibfnamefont {D.}~\bibnamefont
  {{Merritt}}},\ }\bibfield  {title} {\bibinfo {title} {{Dark Matter Dynamics
  and Indirect Detection}},\ }\href {https://doi.org/10.1142/S0217732305017391}
  {\bibfield  {journal} {\bibinfo  {journal} {Modern Physics Letters A}\
  }\textbf {\bibinfo {volume} {20}},\ \bibinfo {pages} {1021} (\bibinfo {year}
  {2005})},\ \Eprint {https://arxiv.org/abs/astro-ph/0504422}
  {arXiv:astro-ph/0504422 [astro-ph]} \BibitemShut {NoStop}%
\bibitem [{\citenamefont {{Merritt}}(2010)}]{2010arXiv1001.3706M}%
  \BibitemOpen
  \bibfield  {author} {\bibinfo {author} {\bibfnamefont {D.}~\bibnamefont
  {{Merritt}}},\ }\bibfield  {title} {\bibinfo {title} {{Dark matter at the
  centers of galaxies}},\ }\href@noop {} {\bibfield  {journal} {\bibinfo
  {journal} {arXiv e-prints}\ ,\ \bibinfo {eid} {arXiv:1001.3706}} (\bibinfo
  {year} {2010})},\ \Eprint {https://arxiv.org/abs/1001.3706} {arXiv:1001.3706
  [astro-ph.CO]} \BibitemShut {NoStop}%
\bibitem [{\citenamefont {{Ellis}}\ \emph {et~al.}(1988)\citenamefont
  {{Ellis}}, \citenamefont {{Flores}}, \citenamefont {{Freese}}, \citenamefont
  {{Ritz}}, \citenamefont {{Seckel}},\ and\ \citenamefont
  {{Silk}}}]{1988PhLB..214..403E}%
  \BibitemOpen
  \bibfield  {author} {\bibinfo {author} {\bibfnamefont {J.}~\bibnamefont
  {{Ellis}}}, \bibinfo {author} {\bibfnamefont {R.~A.}\ \bibnamefont
  {{Flores}}}, \bibinfo {author} {\bibfnamefont {K.}~\bibnamefont {{Freese}}},
  \bibinfo {author} {\bibfnamefont {S.}~\bibnamefont {{Ritz}}}, \bibinfo
  {author} {\bibfnamefont {D.}~\bibnamefont {{Seckel}}},\ and\ \bibinfo
  {author} {\bibfnamefont {J.}~\bibnamefont {{Silk}}},\ }\bibfield  {title}
  {\bibinfo {title} {{Cosmic ray constraints on the annihilations of relic
  particles in the galactic halo}},\ }\href
  {https://doi.org/10.1016/0370-2693(88)91385-8} {\bibfield  {journal}
  {\bibinfo  {journal} {Physics Letters B}\ }\textbf {\bibinfo {volume}
  {214}},\ \bibinfo {pages} {403} (\bibinfo {year} {1988})}\BibitemShut
  {NoStop}%
\bibitem [{\citenamefont {Levkov}\ \emph {et~al.}(2017)\citenamefont {Levkov},
  \citenamefont {Panin},\ and\ \citenamefont {Tkachev}}]{Levkov_2017}%
  \BibitemOpen
  \bibfield  {author} {\bibinfo {author} {\bibfnamefont {D.}~\bibnamefont
  {Levkov}}, \bibinfo {author} {\bibfnamefont {A.}~\bibnamefont {Panin}},\ and\
  \bibinfo {author} {\bibfnamefont {I.}~\bibnamefont {Tkachev}},\ }\bibfield
  {title} {\bibinfo {title} {Relativistic axions from collapsing bose stars},\
  }\bibfield  {journal} {\bibinfo  {journal} {Physical Review Letters}\
  }\textbf {\bibinfo {volume} {118}},\ \href
  {https://doi.org/10.1103/physrevlett.118.011301}
  {10.1103/physrevlett.118.011301} (\bibinfo {year} {2017})\BibitemShut
  {NoStop}%
\bibitem [{\citenamefont {Fornberg}(1987)}]{Fornberg1987}%
  \BibitemOpen
  \bibfield  {author} {\bibinfo {author} {\bibfnamefont {B.}~\bibnamefont
  {Fornberg}},\ }\bibfield  {title} {\bibinfo {title} {The pseudospectral
  method: Comparisons with finite differences for the elastic wave equation},\
  }\href {https://doi.org/10.1190/1.1442319} {\bibfield  {journal} {\bibinfo
  {journal} {GEOPHYSICS}\ }\textbf {\bibinfo {volume} {52}},\ \bibinfo {pages}
  {483} (\bibinfo {year} {1987})},\ \Eprint
  {https://arxiv.org/abs/https://doi.org/10.1190/1.1442319}
  {https://doi.org/10.1190/1.1442319} \BibitemShut {NoStop}%
\bibitem [{\citenamefont {Magaña}\ and\ \citenamefont
  {Matos}(2012)}]{Maga_a_2012}%
  \BibitemOpen
  \bibfield  {author} {\bibinfo {author} {\bibfnamefont {J.}~\bibnamefont
  {Magaña}}\ and\ \bibinfo {author} {\bibfnamefont {T.}~\bibnamefont
  {Matos}},\ }\bibfield  {title} {\bibinfo {title} {A brief review of the
  scalar field dark matter model},\ }\href
  {https://doi.org/10.1088/1742-6596/378/1/012012} {\bibfield  {journal}
  {\bibinfo  {journal} {Journal of Physics: Conference Series}\ }\textbf
  {\bibinfo {volume} {378}},\ \bibinfo {pages} {012012} (\bibinfo {year}
  {2012})}\BibitemShut {NoStop}%
\bibitem [{\citenamefont {Amin}\ and\ \citenamefont {Mocz}(2019)}]{Amin_2019}%
  \BibitemOpen
  \bibfield  {author} {\bibinfo {author} {\bibfnamefont {M.~A.}\ \bibnamefont
  {Amin}}\ and\ \bibinfo {author} {\bibfnamefont {P.}~\bibnamefont {Mocz}},\
  }\bibfield  {title} {\bibinfo {title} {Formation, gravitational clustering,
  and interactions of nonrelativistic solitons in an expanding universe},\
  }\bibfield  {journal} {\bibinfo  {journal} {Physical Review D}\ }\textbf
  {\bibinfo {volume} {100}},\ \href
  {https://doi.org/10.1103/physrevd.100.063507} {10.1103/physrevd.100.063507}
  (\bibinfo {year} {2019})\BibitemShut {NoStop}%
\bibitem [{\citenamefont {Marsh}\ and\ \citenamefont
  {Niemeyer}(2019)}]{Marsh:2018zyw}%
  \BibitemOpen
  \bibfield  {author} {\bibinfo {author} {\bibfnamefont {D.~J.}\ \bibnamefont
  {Marsh}}\ and\ \bibinfo {author} {\bibfnamefont {J.~C.}\ \bibnamefont
  {Niemeyer}},\ }\bibfield  {title} {\bibinfo {title} {{Strong Constraints on
  Fuzzy Dark Matter from Ultrafaint Dwarf Galaxy Eridanus II}},\ }\href
  {https://doi.org/10.1103/PhysRevLett.123.051103} {\bibfield  {journal}
  {\bibinfo  {journal} {Phys. Rev. Lett.}\ }\textbf {\bibinfo {volume} {123}},\
  \bibinfo {pages} {051103} (\bibinfo {year} {2019})},\ \Eprint
  {https://arxiv.org/abs/1810.08543} {arXiv:1810.08543 [astro-ph.CO]}
  \BibitemShut {NoStop}%
\bibitem [{\citenamefont {Bar}\ \emph {et~al.}(2018)\citenamefont {Bar},
  \citenamefont {Blas}, \citenamefont {Blum},\ and\ \citenamefont
  {Sibiryakov}}]{Bar:2018acw}%
  \BibitemOpen
  \bibfield  {author} {\bibinfo {author} {\bibfnamefont {N.}~\bibnamefont
  {Bar}}, \bibinfo {author} {\bibfnamefont {D.}~\bibnamefont {Blas}}, \bibinfo
  {author} {\bibfnamefont {K.}~\bibnamefont {Blum}},\ and\ \bibinfo {author}
  {\bibfnamefont {S.}~\bibnamefont {Sibiryakov}},\ }\bibfield  {title}
  {\bibinfo {title} {{Galactic rotation curves versus ultralight dark matter:
  Implications of the soliton-host halo relation}},\ }\href
  {https://doi.org/10.1103/PhysRevD.98.083027} {\bibfield  {journal} {\bibinfo
  {journal} {Phys. Rev. D}\ }\textbf {\bibinfo {volume} {98}},\ \bibinfo
  {pages} {083027} (\bibinfo {year} {2018})},\ \Eprint
  {https://arxiv.org/abs/1805.00122} {arXiv:1805.00122 [astro-ph.CO]}
  \BibitemShut {NoStop}%
\bibitem [{\citenamefont {{Li}}\ \emph {et~al.}(2020)\citenamefont {{Li}},
  \citenamefont {{Shen}},\ and\ \citenamefont
  {{Schive}}}]{2020ApJ...889...88L}%
  \BibitemOpen
  \bibfield  {author} {\bibinfo {author} {\bibfnamefont {Z.}~\bibnamefont
  {{Li}}}, \bibinfo {author} {\bibfnamefont {J.}~\bibnamefont {{Shen}}},\ and\
  \bibinfo {author} {\bibfnamefont {H.-Y.}\ \bibnamefont {{Schive}}},\
  }\bibfield  {title} {\bibinfo {title} {{Testing the Prediction of Fuzzy Dark
  Matter Theory in the Milky Way Center}},\ }\href
  {https://doi.org/10.3847/1538-4357/ab6598} {\bibfield  {journal} {\bibinfo
  {journal} {\apj}\ }\textbf {\bibinfo {volume} {889}},\ \bibinfo {eid} {88}
  (\bibinfo {year} {2020})},\ \Eprint {https://arxiv.org/abs/2001.00318}
  {arXiv:2001.00318 [astro-ph.GA]} \BibitemShut {NoStop}%
\bibitem [{\citenamefont {Hook}\ \emph {et~al.}(2018)\citenamefont {Hook},
  \citenamefont {Kahn}, \citenamefont {Safdi},\ and\ \citenamefont
  {Sun}}]{Hook:2018iia}%
  \BibitemOpen
  \bibfield  {author} {\bibinfo {author} {\bibfnamefont {A.}~\bibnamefont
  {Hook}}, \bibinfo {author} {\bibfnamefont {Y.}~\bibnamefont {Kahn}}, \bibinfo
  {author} {\bibfnamefont {B.~R.}\ \bibnamefont {Safdi}},\ and\ \bibinfo
  {author} {\bibfnamefont {Z.}~\bibnamefont {Sun}},\ }\bibfield  {title}
  {\bibinfo {title} {{Radio Signals from Axion Dark Matter Conversion in
  Neutron Star Magnetospheres}},\ }\href
  {https://doi.org/10.1103/PhysRevLett.121.241102} {\bibfield  {journal}
  {\bibinfo  {journal} {Phys. Rev. Lett.}\ }\textbf {\bibinfo {volume} {121}},\
  \bibinfo {pages} {241102} (\bibinfo {year} {2018})},\ \Eprint
  {https://arxiv.org/abs/1804.03145} {arXiv:1804.03145 [hep-ph]} \BibitemShut
  {NoStop}%
\bibitem [{\citenamefont {Jackson~Kimball}\ \emph {et~al.}(2018)\citenamefont
  {Jackson~Kimball}, \citenamefont {Budker}, \citenamefont {Eby}, \citenamefont
  {Pospelov}, \citenamefont {Pustelny}, \citenamefont {Scholtes}, \citenamefont
  {Stadnik}, \citenamefont {Weis},\ and\ \citenamefont
  {Wickenbrock}}]{JacksonKimball:2017qgk}%
  \BibitemOpen
  \bibfield  {author} {\bibinfo {author} {\bibfnamefont {D.~F.}\ \bibnamefont
  {Jackson~Kimball}}, \bibinfo {author} {\bibfnamefont {D.}~\bibnamefont
  {Budker}}, \bibinfo {author} {\bibfnamefont {J.}~\bibnamefont {Eby}},
  \bibinfo {author} {\bibfnamefont {M.}~\bibnamefont {Pospelov}}, \bibinfo
  {author} {\bibfnamefont {S.}~\bibnamefont {Pustelny}}, \bibinfo {author}
  {\bibfnamefont {T.}~\bibnamefont {Scholtes}}, \bibinfo {author}
  {\bibfnamefont {Y.}~\bibnamefont {Stadnik}}, \bibinfo {author} {\bibfnamefont
  {A.}~\bibnamefont {Weis}},\ and\ \bibinfo {author} {\bibfnamefont
  {A.}~\bibnamefont {Wickenbrock}},\ }\bibfield  {title} {\bibinfo {title}
  {{Searching for axion stars and Q-balls with a terrestrial magnetometer
  network}},\ }\href {https://doi.org/10.1103/PhysRevD.97.043002} {\bibfield
  {journal} {\bibinfo  {journal} {Phys. Rev. D}\ }\textbf {\bibinfo {volume}
  {97}},\ \bibinfo {pages} {043002} (\bibinfo {year} {2018})},\ \Eprint
  {https://arxiv.org/abs/1710.04323} {arXiv:1710.04323 [physics.atom-ph]}
  \BibitemShut {NoStop}%
\bibitem [{\citenamefont {Giudice}\ \emph {et~al.}(2016)\citenamefont
  {Giudice}, \citenamefont {McCullough},\ and\ \citenamefont
  {Urbano}}]{Giudice:2016zpa}%
  \BibitemOpen
  \bibfield  {author} {\bibinfo {author} {\bibfnamefont {G.~F.}\ \bibnamefont
  {Giudice}}, \bibinfo {author} {\bibfnamefont {M.}~\bibnamefont
  {McCullough}},\ and\ \bibinfo {author} {\bibfnamefont {A.}~\bibnamefont
  {Urbano}},\ }\bibfield  {title} {\bibinfo {title} {{Hunting for Dark
  Particles with Gravitational Waves}},\ }\href
  {https://doi.org/10.1088/1475-7516/2016/10/001} {\bibfield  {journal}
  {\bibinfo  {journal} {JCAP}\ }\textbf {\bibinfo {volume} {10}},\ \bibinfo
  {pages} {001}},\ \Eprint {https://arxiv.org/abs/1605.01209} {arXiv:1605.01209
  [hep-ph]} \BibitemShut {NoStop}%
\bibitem [{\citenamefont {Chavanis}\ and\ \citenamefont
  {Delfini}(2011)}]{Chavanis:2011zm}%
  \BibitemOpen
  \bibfield  {author} {\bibinfo {author} {\bibfnamefont {P.}~\bibnamefont
  {Chavanis}}\ and\ \bibinfo {author} {\bibfnamefont {L.}~\bibnamefont
  {Delfini}},\ }\bibfield  {title} {\bibinfo {title} {{Mass-radius relation of
  Newtonian self-gravitating Bose-Einstein condensates with short-range
  interactions: II. Numerical results}},\ }\href
  {https://doi.org/10.1103/PhysRevD.84.043532} {\bibfield  {journal} {\bibinfo
  {journal} {Phys. Rev. D}\ }\textbf {\bibinfo {volume} {84}},\ \bibinfo
  {pages} {043532} (\bibinfo {year} {2011})},\ \Eprint
  {https://arxiv.org/abs/1103.2054} {arXiv:1103.2054 [astro-ph.CO]}
  \BibitemShut {NoStop}%
\bibitem [{\citenamefont {Eby}\ \emph {et~al.}(2016)\citenamefont {Eby},
  \citenamefont {Kouvaris}, \citenamefont {Nielsen},\ and\ \citenamefont
  {Wijewardhana}}]{Eby:2015hsq}%
  \BibitemOpen
  \bibfield  {author} {\bibinfo {author} {\bibfnamefont {J.}~\bibnamefont
  {Eby}}, \bibinfo {author} {\bibfnamefont {C.}~\bibnamefont {Kouvaris}},
  \bibinfo {author} {\bibfnamefont {N.~G.}\ \bibnamefont {Nielsen}},\ and\
  \bibinfo {author} {\bibfnamefont {L.}~\bibnamefont {Wijewardhana}},\
  }\bibfield  {title} {\bibinfo {title} {{Boson Stars from Self-Interacting
  Dark Matter}},\ }\href {https://doi.org/10.1007/JHEP02(2016)028} {\bibfield
  {journal} {\bibinfo  {journal} {JHEP}\ }\textbf {\bibinfo {volume} {02}},\
  \bibinfo {pages} {028}},\ \Eprint {https://arxiv.org/abs/1511.04474}
  {arXiv:1511.04474 [hep-ph]} \BibitemShut {NoStop}%
\bibitem [{\citenamefont {Khlopov}\ \emph {et~al.}(1985)\citenamefont
  {Khlopov}, \citenamefont {Malomed},\ and\ \citenamefont
  {Zeldovich}}]{khlopov_scalar}%
  \BibitemOpen
  \bibfield  {author} {\bibinfo {author} {\bibfnamefont {M.}~\bibnamefont
  {Khlopov}}, \bibinfo {author} {\bibfnamefont {B.}~\bibnamefont {Malomed}},\
  and\ \bibinfo {author} {\bibfnamefont {I.}~\bibnamefont {Zeldovich}},\
  }\bibfield  {title} {\bibinfo {title} {{Gravitational instability of scalar
  fields and formation of primordial black holes}},\ }\href@noop {} {\bibfield
  {journal} {\bibinfo  {journal} {\mnras}\ }\textbf {\bibinfo {volume} {215}},\
  \bibinfo {pages} {575} (\bibinfo {year} {1985})}\BibitemShut {NoStop}%
\bibitem [{\citenamefont {Kirkpatrick}\ \emph {et~al.}(2020)\citenamefont
  {Kirkpatrick}, \citenamefont {Mirasola},\ and\ \citenamefont
  {Prescod-Weinstein}}]{Kirkpatrick:2020fwd}%
  \BibitemOpen
  \bibfield  {author} {\bibinfo {author} {\bibfnamefont {K.}~\bibnamefont
  {Kirkpatrick}}, \bibinfo {author} {\bibfnamefont {A.~E.}\ \bibnamefont
  {Mirasola}},\ and\ \bibinfo {author} {\bibfnamefont {C.}~\bibnamefont
  {Prescod-Weinstein}},\ }\bibfield  {title} {\bibinfo {title} {{Relaxation
  times for Bose-Einstein condensation in axion miniclusters}},\ }\href
  {https://doi.org/10.1103/PhysRevD.102.103012} {\bibfield  {journal} {\bibinfo
   {journal} {Phys. Rev. D}\ }\textbf {\bibinfo {volume} {102}},\ \bibinfo
  {pages} {103012} (\bibinfo {year} {2020})},\ \Eprint
  {https://arxiv.org/abs/2007.07438} {arXiv:2007.07438 [hep-ph]} \BibitemShut
  {NoStop}%
\bibitem [{\citenamefont {{Chavanis}}(2011)}]{2011PhRvD..84d3531C}%
  \BibitemOpen
  \bibfield  {author} {\bibinfo {author} {\bibfnamefont {P.-H.}\ \bibnamefont
  {{Chavanis}}},\ }\bibfield  {title} {\bibinfo {title} {{Mass-radius relation
  of Newtonian self-gravitating Bose-Einstein condensates with short-range
  interactions. I. Analytical results}},\ }\href
  {https://doi.org/10.1103/PhysRevD.84.043531} {\bibfield  {journal} {\bibinfo
  {journal} {\prd}\ }\textbf {\bibinfo {volume} {84}},\ \bibinfo {eid} {043531}
  (\bibinfo {year} {2011})},\ \Eprint {https://arxiv.org/abs/1103.2050}
  {arXiv:1103.2050} \BibitemShut {NoStop}%
\bibitem [{\citenamefont {Chavanis}(2016)}]{Chavanis:2016dab}%
  \BibitemOpen
  \bibfield  {author} {\bibinfo {author} {\bibfnamefont {P.-H.}\ \bibnamefont
  {Chavanis}},\ }\bibfield  {title} {\bibinfo {title} {{Collapse of a
  self-gravitating Bose-Einstein condensate with attractive
  self-interaction}},\ }\href {https://doi.org/10.1103/PhysRevD.94.083007}
  {\bibfield  {journal} {\bibinfo  {journal} {Phys. Rev. D}\ }\textbf {\bibinfo
  {volume} {94}},\ \bibinfo {pages} {083007} (\bibinfo {year} {2016})},\
  \Eprint {https://arxiv.org/abs/1604.05904} {arXiv:1604.05904 [astro-ph.CO]}
  \BibitemShut {NoStop}%
\bibitem [{\citenamefont {Cembranos}\ \emph {et~al.}(2018)\citenamefont
  {Cembranos}, \citenamefont {Maroto}, \citenamefont {Núñez~Jareño},\ and\
  \citenamefont {Villarrubia-Rojo}}]{Cembranos_2018}%
  \BibitemOpen
  \bibfield  {author} {\bibinfo {author} {\bibfnamefont {J.~A.~R.}\
  \bibnamefont {Cembranos}}, \bibinfo {author} {\bibfnamefont {A.~L.}\
  \bibnamefont {Maroto}}, \bibinfo {author} {\bibfnamefont {S.~J.}\
  \bibnamefont {Núñez~Jareño}},\ and\ \bibinfo {author} {\bibfnamefont
  {H.}~\bibnamefont {Villarrubia-Rojo}},\ }\bibfield  {title} {\bibinfo {title}
  {Constraints on anharmonic corrections of fuzzy dark matter},\ }\bibfield
  {journal} {\bibinfo  {journal} {Journal of High Energy Physics}\ }\textbf
  {\bibinfo {volume} {2018}},\ \href {https://doi.org/10.1007/jhep08(2018)073}
  {10.1007/jhep08(2018)073} (\bibinfo {year} {2018})\BibitemShut {NoStop}%
\bibitem [{\citenamefont {Veltmaat}\ \emph {et~al.}(2018)\citenamefont
  {Veltmaat}, \citenamefont {Niemeyer},\ and\ \citenamefont
  {Schwabe}}]{Veltmaat:2018dfz}%
  \BibitemOpen
  \bibfield  {author} {\bibinfo {author} {\bibfnamefont {J.}~\bibnamefont
  {Veltmaat}}, \bibinfo {author} {\bibfnamefont {J.~C.}\ \bibnamefont
  {Niemeyer}},\ and\ \bibinfo {author} {\bibfnamefont {B.}~\bibnamefont
  {Schwabe}},\ }\bibfield  {title} {\bibinfo {title} {{Formation and structure
  of ultralight bosonic dark matter halos}},\ }\href
  {https://doi.org/10.1103/PhysRevD.98.043509} {\bibfield  {journal} {\bibinfo
  {journal} {Phys. Rev. D}\ }\textbf {\bibinfo {volume} {98}},\ \bibinfo
  {pages} {043509} (\bibinfo {year} {2018})},\ \Eprint
  {https://arxiv.org/abs/1804.09647} {arXiv:1804.09647 [astro-ph.CO]}
  \BibitemShut {NoStop}%
\bibitem [{\citenamefont {Chavanis}(2012)}]{Chavanis:2011uv}%
  \BibitemOpen
  \bibfield  {author} {\bibinfo {author} {\bibfnamefont {P.-H.}\ \bibnamefont
  {Chavanis}},\ }\bibfield  {title} {\bibinfo {title} {{Growth of perturbations
  in an expanding universe with Bose-Einstein condensate dark matter}},\ }\href
  {https://doi.org/10.1051/0004-6361/201116905} {\bibfield  {journal} {\bibinfo
   {journal} {Astron. Astrophys.}\ }\textbf {\bibinfo {volume} {537}},\
  \bibinfo {pages} {A127} (\bibinfo {year} {2012})},\ \Eprint
  {https://arxiv.org/abs/1103.2698} {arXiv:1103.2698 [astro-ph.CO]}
  \BibitemShut {NoStop}%
\bibitem [{\citenamefont {Desjacques}\ \emph {et~al.}(2018)\citenamefont
  {Desjacques}, \citenamefont {Kehagias},\ and\ \citenamefont
  {Riotto}}]{Desjacques_2018}%
  \BibitemOpen
  \bibfield  {author} {\bibinfo {author} {\bibfnamefont {V.}~\bibnamefont
  {Desjacques}}, \bibinfo {author} {\bibfnamefont {A.}~\bibnamefont
  {Kehagias}},\ and\ \bibinfo {author} {\bibfnamefont {A.}~\bibnamefont
  {Riotto}},\ }\bibfield  {title} {\bibinfo {title} {Impact of ultralight axion
  self-interactions on the large scale structure of the universe},\ }\bibfield
  {journal} {\bibinfo  {journal} {Physical Review D}\ }\textbf {\bibinfo
  {volume} {97}},\ \href {https://doi.org/10.1103/physrevd.97.023529}
  {10.1103/physrevd.97.023529} (\bibinfo {year} {2018})\BibitemShut {NoStop}%
\bibitem [{\citenamefont {Niemeyer}\ and\ \citenamefont
  {Easther}(2020)}]{Niemeyer_2020}%
  \BibitemOpen
  \bibfield  {author} {\bibinfo {author} {\bibfnamefont {J.~C.}\ \bibnamefont
  {Niemeyer}}\ and\ \bibinfo {author} {\bibfnamefont {R.}~\bibnamefont
  {Easther}},\ }\bibfield  {title} {\bibinfo {title} {Inflaton clusters and
  inflaton stars},\ }\href {https://doi.org/10.1088/1475-7516/2020/07/030}
  {\bibfield  {journal} {\bibinfo  {journal} {Journal of Cosmology and
  Astroparticle Physics}\ }\textbf {\bibinfo {volume} {2020}}\bibinfo  {number}
  { (07)},\ \bibinfo {pages} {030–030}}\BibitemShut {NoStop}%
\bibitem [{\citenamefont {Du}\ \emph {et~al.}(2017)\citenamefont {Du},
  \citenamefont {Behrens}, \citenamefont {Niemeyer},\ and\ \citenamefont
  {Schwabe}}]{Du:2016aik}%
  \BibitemOpen
\bibfield  {number} {  }\bibfield  {author} {\bibinfo {author} {\bibfnamefont
  {X.}~\bibnamefont {Du}}, \bibinfo {author} {\bibfnamefont {C.}~\bibnamefont
  {Behrens}}, \bibinfo {author} {\bibfnamefont {J.~C.}\ \bibnamefont
  {Niemeyer}},\ and\ \bibinfo {author} {\bibfnamefont {B.}~\bibnamefont
  {Schwabe}},\ }\bibfield  {title} {\bibinfo {title} {{Core-halo mass relation
  of ultralight axion dark matter from merger history}},\ }\href
  {https://doi.org/10.1103/PhysRevD.95.043519} {\bibfield  {journal} {\bibinfo
  {journal} {Phys. Rev. D}\ }\textbf {\bibinfo {volume} {95}},\ \bibinfo
  {pages} {043519} (\bibinfo {year} {2017})},\ \Eprint
  {https://arxiv.org/abs/1609.09414} {arXiv:1609.09414 [astro-ph.GA]}
  \BibitemShut {NoStop}%
\bibitem [{\citenamefont {Musoke}\ \emph {et~al.}(2020)\citenamefont {Musoke},
  \citenamefont {Hotchkiss},\ and\ \citenamefont {Easther}}]{Musoke:2019ima}%
  \BibitemOpen
  \bibfield  {author} {\bibinfo {author} {\bibfnamefont {N.}~\bibnamefont
  {Musoke}}, \bibinfo {author} {\bibfnamefont {S.}~\bibnamefont {Hotchkiss}},\
  and\ \bibinfo {author} {\bibfnamefont {R.}~\bibnamefont {Easther}},\
  }\bibfield  {title} {\bibinfo {title} {{Lighting the Dark: Evolution of the
  Postinflationary Universe}},\ }\href
  {https://doi.org/10.1103/PhysRevLett.124.061301} {\bibfield  {journal}
  {\bibinfo  {journal} {Phys. Rev. Lett.}\ }\textbf {\bibinfo {volume} {124}},\
  \bibinfo {pages} {061301} (\bibinfo {year} {2020})},\ \Eprint
  {https://arxiv.org/abs/1909.11678} {arXiv:1909.11678 [astro-ph.CO]}
  \BibitemShut {NoStop}%
\bibitem [{\citenamefont {Clough}\ \emph {et~al.}(2018)\citenamefont {Clough},
  \citenamefont {Dietrich},\ and\ \citenamefont {Niemeyer}}]{Clough:2018exo}%
  \BibitemOpen
  \bibfield  {author} {\bibinfo {author} {\bibfnamefont {K.}~\bibnamefont
  {Clough}}, \bibinfo {author} {\bibfnamefont {T.}~\bibnamefont {Dietrich}},\
  and\ \bibinfo {author} {\bibfnamefont {J.~C.}\ \bibnamefont {Niemeyer}},\
  }\bibfield  {title} {\bibinfo {title} {{Axion star collisions with black
  holes and neutron stars in full 3D numerical relativity}},\ }\href
  {https://doi.org/10.1103/PhysRevD.98.083020} {\bibfield  {journal} {\bibinfo
  {journal} {Phys. Rev. D}\ }\textbf {\bibinfo {volume} {98}},\ \bibinfo
  {pages} {083020} (\bibinfo {year} {2018})},\ \Eprint
  {https://arxiv.org/abs/1808.04668} {arXiv:1808.04668 [gr-qc]} \BibitemShut
  {NoStop}%
\bibitem [{cud()}]{cuda_gpu}%
  \BibitemOpen
  \href {https://developer.nvidia.com/cufft} {\emph {\bibinfo {title} {cuFFT |
  NVIDIA Developer}}}\BibitemShut {NoStop}%
\bibitem [{\citenamefont {Du}\ \emph {et~al.}(2018)\citenamefont {Du},
  \citenamefont {Schwabe}, \citenamefont {Niemeyer},\ and\ \citenamefont
  {Bürger}}]{Du:2018qor}%
  \BibitemOpen
  \bibfield  {author} {\bibinfo {author} {\bibfnamefont {X.}~\bibnamefont
  {Du}}, \bibinfo {author} {\bibfnamefont {B.}~\bibnamefont {Schwabe}},
  \bibinfo {author} {\bibfnamefont {J.~C.}\ \bibnamefont {Niemeyer}},\ and\
  \bibinfo {author} {\bibfnamefont {D.}~\bibnamefont {Bürger}},\ }\bibfield
  {title} {\bibinfo {title} {{Tidal disruption of fuzzy dark matter subhalo
  cores}},\ }\href {https://doi.org/10.1103/PhysRevD.97.063507} {\bibfield
  {journal} {\bibinfo  {journal} {Phys. Rev. D}\ }\textbf {\bibinfo {volume}
  {97}},\ \bibinfo {pages} {063507} (\bibinfo {year} {2018})},\ \Eprint
  {https://arxiv.org/abs/1801.04864} {arXiv:1801.04864 [astro-ph.GA]}
  \BibitemShut {NoStop}%
\bibitem [{\citenamefont {McLachlan}(1995)}]{McLachlan:1995}%
  \BibitemOpen
  \bibfield  {author} {\bibinfo {author} {\bibfnamefont {R.~I.}\ \bibnamefont
  {McLachlan}},\ }\bibfield  {title} {\bibinfo {title} {On the numerical
  integration of ordinary differential equations by symmetric composition
  methods},\ }\href {https://doi.org/10.1137/0916010} {\bibfield  {journal}
  {\bibinfo  {journal} {SIAM Journal on Scientific Computing}\ }\textbf
  {\bibinfo {volume} {16}},\ \bibinfo {pages} {151} (\bibinfo {year}
  {1995})}\BibitemShut {NoStop}%
\bibitem [{\citenamefont {Chin}(2007)}]{chin2007forward}%
  \BibitemOpen
  \bibfield  {author} {\bibinfo {author} {\bibfnamefont {S.~A.}\ \bibnamefont
  {Chin}},\ }\href@noop {} {\bibinfo {title} {Forward and non-forward
  symplectic integrators in solving classical dynamics problems}} (\bibinfo
  {year} {2007}),\ \Eprint {https://arxiv.org/abs/0704.3273} {arXiv:0704.3273
  [physics.comp-ph]} \BibitemShut {NoStop}%
\bibitem [{\citenamefont {Woo}\ and\ \citenamefont
  {Chiueh}(2009)}]{Woo:2008nn}%
  \BibitemOpen
  \bibfield  {author} {\bibinfo {author} {\bibfnamefont {T.-P.}\ \bibnamefont
  {Woo}}\ and\ \bibinfo {author} {\bibfnamefont {T.}~\bibnamefont {Chiueh}},\
  }\bibfield  {title} {\bibinfo {title} {{High-Resolution Simulation on
  Structure Formation with Extremely Light Bosonic Dark Matter}},\ }\href
  {https://doi.org/10.1088/0004-637X/697/1/850} {\bibfield  {journal} {\bibinfo
   {journal} {Astrophys. J.}\ }\textbf {\bibinfo {volume} {697}},\ \bibinfo
  {pages} {850} (\bibinfo {year} {2009})},\ \Eprint
  {https://arxiv.org/abs/0806.0232} {arXiv:0806.0232 [astro-ph]} \BibitemShut
  {NoStop}%
\bibitem [{\citenamefont {Kling}\ and\ \citenamefont
  {Rajaraman}(2017)}]{Kling:2017mif}%
  \BibitemOpen
  \bibfield  {author} {\bibinfo {author} {\bibfnamefont {F.}~\bibnamefont
  {Kling}}\ and\ \bibinfo {author} {\bibfnamefont {A.}~\bibnamefont
  {Rajaraman}},\ }\bibfield  {title} {\bibinfo {title} {{Towards an Analytic
  Construction of the Wavefunction of Boson Stars}},\ }\href
  {https://doi.org/10.1103/PhysRevD.96.044039} {\bibfield  {journal} {\bibinfo
  {journal} {Phys. Rev. D}\ }\textbf {\bibinfo {volume} {96}},\ \bibinfo
  {pages} {044039} (\bibinfo {year} {2017})},\ \Eprint
  {https://arxiv.org/abs/1706.04272} {arXiv:1706.04272 [hep-th]} \BibitemShut
  {NoStop}%
\bibitem [{\citenamefont {Kling}\ and\ \citenamefont
  {Rajaraman}(2018)}]{Kling:2017hjm}%
  \BibitemOpen
  \bibfield  {author} {\bibinfo {author} {\bibfnamefont {F.}~\bibnamefont
  {Kling}}\ and\ \bibinfo {author} {\bibfnamefont {A.}~\bibnamefont
  {Rajaraman}},\ }\bibfield  {title} {\bibinfo {title} {{Profiles of boson
  stars with self-interactions}},\ }\href
  {https://doi.org/10.1103/PhysRevD.97.063012} {\bibfield  {journal} {\bibinfo
  {journal} {Phys. Rev. D}\ }\textbf {\bibinfo {volume} {97}},\ \bibinfo
  {pages} {063012} (\bibinfo {year} {2018})},\ \Eprint
  {https://arxiv.org/abs/1712.06539} {arXiv:1712.06539 [hep-ph]} \BibitemShut
  {NoStop}%
\end{thebibliography}%

\end{document}